# Probabilities of collisions of bodies ejected from forming Earth with the terrestrial planets

S.I. Ipatov [*]

*Vernadsky Institute of Geochemistry and Analytical Chemistry of RAS, Moscow, Russia*

ARTICLE INFO

ABSTRACT

During formation of the Earth and at the stage of the Late Heavy Bombardment, some bodies collided with the Earth. Such collisions caused ejection of material from the Earth. The motion of bodies ejected from the Earth was studied, and the probabilities of collisions of such bodies with the present terrestrial planets were calculated. The dependences of these probabilities on velocities, angles and points of ejection of bodies were studied. These dependences can be used in the models with different distributions of ejected material. On average, about a half and less than 10 % of initial ejected bodies remained moving in elliptical orbits in the Solar System after 10 and 100 Myr, respectively. A few ejected bodies collided with planets after 250 Myr. As dynamical lifetimes of bodies ejected from the Earth can reach hundreds of million years, a few percent of bodies ejected at the Chicxulub and Popigai events about 36–65 Myr ago can still move in the zone of the terrestrial planets and have small chances to collide with planets, including the Earth. The fraction of ejected bodies that collided with the Earth was greater for smaller ejection velocity. The fractions of bodies delivered to the Earth and Venus probably did not differ much for these planets and were about 0.2–0.3 each. Such obtained results testify in favour of that the upper layers of the Earth and Venus can contain similar material. The fractions of bodies ejected from the Earth that collided with Mercury and Mars did not exceed 0.08 and 0.025, respectively. The fractions of bodies collided with Jupiter were of the order of 0.001. In most calculations the fraction of bodies collided with the Sun was between 0.2 and 0.5. Depending on parameters of ejection, the fraction of bodies ejected into hyperbolic orbits could vary from 0 to 1. Small fractions of material ejected from the Earth can be found on other terrestrial planets and Jupiter, as the ejected bodies could collide with these planets. Bodies ejected from the Earth could deliver organic material to other celestial objects, e.g. to Mars.

## 1. Introduction

This paper is devoted to the studies of the motion of bodies ejected from the Earth and to the calculations of the probabilities of collisions of such ejected bodies with different planets and the Sun. The studies were made for different ejection velocities and angles of bodies started from several points on the Earth's surface. In Section 1 I discuss the previous studies of collisions of migrated bodies with the Earth from different distances from the Sun (Section 1.1), the ejection of material from the Earth after collisions of impactors with the Earth (Section 1.2), and the evolution of orbits of bodies ejected from the terrestrial planets (Section 1.3). The problems considered in the paper are shortly summarized in Section 1.4.

### 1.1. Collisions of migrated bodies with the Earth

Growing planets were bombarded by planetesimals during planet formation. Together with analytical estimates (e.g., Lissauer, 1993; Safronov, 1972; Wetherill, 1980), there are a large number of papers based on the results of numerical simulations of the formation of the terrestrial planets. The results of calculations of disks of gravitating bodies which coagulate under collisions in the feeding zone of the terrestrial planets were presented e.g. by Carter and Stewart (2022), Chambers (2001, 2013), Chambers and Wetherill (1998), Clement et al. (2018, 2019, 2021), Ipatov (1987, 1993, 2000), Izidoro et al. (2014), Joiret et al. (2023), Lykawka and Ito (2017, 2024), Marov and Ipatov (2023), Morbidelli et al. (2012a), Morishima et al. (2010), Nesvorný et al. (2021b), O'Brien et al. (2014), Raymond et al. (2004, 2006, 2009), Wetherill (1980), Woo et al. (2022, 2024). The obtained results include the studies of times of accumulation of planets, variations of mean eccentricities of planetesimals with time, mixing of planetesimals initially located at different distances from the Sun. Some of these results are discussed below in the main text of the paper.

Masses of the largest bodies collided with the Earth were discussed in the models of formation of the Moon. In the Giant Impact Model of Moon formation (e.g., Barr, 2016; Cameron and Ward, 1976; Canup, 2004, 2012; Canup and Asphaug, 2001; Canup et al., 2013; Cuk and Stewart, 2012; Cuk et al., 2016; Hartmann and Davis, 1975) the mass of the impactor was about the mass of Mars. Multi-impactor models of formation of the Moon (e.g., Citron et al., 2014; Gorkavyi, 2023; Ringwood, 1989; Rufu and Aharonson, 2015; Rufu et al., 2017, 2021; Svetsov et al., 2012) allow smaller masses of impactors than the Giant Impact Model. Rufu et al. (2017) considered about 20 impactors with masses from $0.01 m_E$ to $0.09 m_E$, where $m_E$ is the Earth mass. In different models, the timing of a collision of forming Earth with a Mars-sized object was about 20–100 (mainly 40–60) Myr (see e.g. O'Brien et al., 2014; Nesvorný et al., 2022, 2023).

A lot of bodies could collide with the terrestrial planets and the Moon after the Earth formation. This period, called the Late Heavy Bombardment (LHB), is considered to be 4.5–3.5 Gyr ago in (Marchi et al., 2014), 4.2–3.5 Gyr ago in (Bottke and Norman, 2017), 4.25–3.87 Gyr in (Michael et al., 2018), and 4.5–3.8 Gyr ago in (Liu et al., 2023). Four main sources of the Late Heavy Bombardment were discussed in Section 3 in (Bottke and Norman, 2017). According to (Bottke and Norman, 2017), they include (1) a population of planetesimals left over from accretion of the terrestrial planets; (2) a population of objects that escaped from the young main asteroid belt; (3) the hypothetical population of objects that formed near/in the asteroid belt or beyond Saturn's orbit that was left on very high eccentricity, Earth-crossing orbits by early giant planet migration while the solar nebula still existed; (4) the bodies (initially located mainly at distances of about 20–30 AU from the Sun) that changed their orbits due to the migration of the giant planets after the solar nebula had dissipated (e.g. for the Nice model). Nesvorný et al. (2022, 2023) found that the terrestrial-zone planetesimals were the dominant source of lunar impactors for age $T > 3.5$ Gyr, and that asteroids were the dominant source of impactors for $T < 3.5$ Gyr. The E-belt, representing the inner extension of the main asteroid belt at 1.7–2.1 AU, was hypothesized by Bottke et al. (2012) to be an important source of terrestrial and lunar impactors. In their calculations, this region was destabilized by late giant planet migration (the Nice model).

Liu et al. (2023) concluded that the U—Pb age distribution of Apollo samples shows significant contributions of impact melt older than 4.1 Gyr with two peaks at 4.15 and 4.3 Gyr. Morbidelli et al. (2012b) supposed (see Fig. 3 in their paper) that the number of bodies collided with the Moon per unit of time could have a local peak 4.1 Gyr ago. The impact flux at the beginning of this weaker cataclysm was 5–10 times higher than the immediately preceding period. According to Morbidelli et al. (2012b), the lunar bombardment from 4.1 Gyr included the cataclysm caused by the destabilization of the main asteroid belt (~15–25 % of cataclysm impactors) and the E-belt (~75–85 % of cataclysm impactors). Such bombardment accounted for 25 % of the total bombardment suffered by the Moon since it formed. Morbidelli et al. (2012b) considered that bodies from the E-belt could survive during several hundreds of Myr after the destabilization caused by migration of the giant planets. They noted that the bombardment of the terrestrial planets before the lunar cataclysm was presumably caused by remnant planetesimals from the original disk that formed the planets. Computer simulations by Pokorny and Vokrouhlicky (2013) showed that 15 % of bodies of the E-belt collided with the Earth and 16 % - with Mars. Nesvorný et al. (2022, 2023) studied the number of collisions of leftover (from the distance 0.5–1.5 AU from the Sun) planetesimals with a diameter $d > 20$ km with the terrestrial planets and the Moon at different times. Nesvorný et al. (2022) presented the formulas for asteroid and comet impact fluxes to the Earth (depending on $d$ and time $t$ after first solids). They concluded that cometary flux was never large enough in the whole history of the inner Solar System, and the Earth receives about 20 times more impacts than the Moon. Nesvorný et al. (2023) presented plots of cumulative number of impacts of $d > 20$ km (or $d > 10$ km) bodies with the Moon, Earth, Venus and Mars at different times $t$. The plots were for leftover planetesimals, asteroids, and comets. For example, Nesvorný et al. (2023) concluded that about 500 leftover planetesimals with $d > 20$ km are expected to impact the Moon since its formation. Other publications devoted to estimates of the amounts of material delivered from different regions of the Solar System to the Earth are discussed below in Section 4.

According to Bottke et al. (2010), the largest late terrestrial impactors, at 2500 to 3000 km in diameter, potentially modified Earth's obliquity by ~10°, whereas those for the Moon, at ~250 to 300 km, may have delivered water to its mantle. Marchi et al. (2014) supposed that during Earth's late accretion it could collide with bodies up to 4000 km in diameter before 4.15 Gyr. Maximum sizes of later impactors were smaller.

Results presented in Section 1.1 show that some bodies that were initially located in different parts of the Solar System (from the zone of the terrestrial planets to the zone of the giant planets) could collide with almost formed Earth. Scientists estimate that the Solar System is 4.6 Gyr old. Summarizing the above data one can conclude that the large bombardment of the Earth by bodies was during the first billion years after formation of the Sun, and diameters of impactors could exceed a few thousands km during the first 0.5 Gyr. After this first billion years, some bodies with diameters $d > 10$ km also could collide with the Earth. The fraction of bodies originating beyond the orbit of Mars that could collide with the Earth generally increased with time, and composition of impactors on average changed with time.

### 1.2. Ejection of material after a collision

Characteristic velocities of collisions of impactors with the Earth were typically greater for greater initial semi-major axes $a_o$ of orbits of bodies at $a_o > 1$ AU. Such characteristic velocities were about 13–15 km/s for $0.9 \le a_o \le 1.1$ AU, 13–19 km/s for bodies from other parts of the feeding zone of the terrestrial planets (Marov and Ipatov, 2021), 20 km/s for large near-Earth asteroids (Nesvorný et al., 2021b), 21–23.5 km/s for asteroids with $1.6 \le a_o \le 3.3$ AU (Nesvorný et al., 2017), 23–26 km/s for $a_o$ about 15 AU (Ipatov, 2021). However, for some bodies the values of velocities can be outside of the above intervals and could exceed 40 km/s. Velocities of meteoroids entering the Earth's atmosphere can range from about 11 km/s to above 73 km/s (Drolshagen et al., 2020). Armstrong et al. (2002) noted that ejection velocities $v_{ej}$ are mainly less than $0.85 v_{imp}$, where $v_{imp}$ is the impact velocity. Note that the values of $0.85 v_{imp}$ are equal to 12.75, 16.15, and 22.1 km/s at $v_{imp}$ equal to 15, 19, and 26 km/s. Therefore, we can expect that ejection velocities $v_{ej}$ were mainly less than 13, 16, and 22 km/s for bodies that collided with the Earth and came from the feeding zone of the Earth, the feeding zone of the terrestrial planets, and the zone of the giant planets, respectively. So, ejection velocities were greater for the bodies that came from the zone of the giant planets (one of possible sources of the Late Heavy Bombardment) than at the accumulation of the Earth from its feeding zone. Even greater collision and ejection velocities could be at collisions of long-period comets with the Earth.

The mass $m_{ej}$ of material ejected after a collision is smaller at a greater velocity $v_{ej}$ of ejection. According to formula (7.12.3) from (Melosh, 1989), $m_{ej} = \rho\ R^3\ C_{ej}\ v_{ej}^s\ (g\ R)^{0.5v}$, where $R$ is the radius of the crater, $g$ is the gravitational constant, $\rho$ is the density of the material, $C_{ej}$ is a coefficient, and $v$ equals to 1.7 for water and to 1.2 for sand. According to (Svetsov, 2011), $m_{ej}$ is proportional to $v_{ej}^{-1.65}$. Therefore, more material leaves the Earth with smaller velocities (but greater than the parabolic velocity that equals to 11.2 km/s). At values $v_{ej}$ equal to 11.2 (the escape velocity) and 18 (the upper limit for most values of $v_{ej}$), the ratio of values of $v_{ej}^{-1.65}$ is equal to 2.2. Integrating $v_{ej}^{-1.65}$ one gets $v_{ej}^{-0.65}$. Considering interval of velocities from 11.22 to 18 km/s for the equation $11.22^{-0.65} - x^{-0.65} = x^{-0.65} \cdot 18^{-0.65}$ we have $x \approx 14$ km/s. These estimates show that in this case a half of material is ejected with $v_{ej} \geq 14$ km/s, and it is needed to consider not only velocities very close to the parabolic velocity.

According to fig. 12.3 from (Melosh, 1989), for vertical impacts, the mass of material ejected from the Earth was approximately 0.01 and 0.1 of the mass of the impactor at impact velocities $v_{imp}$ of 30 and 45 km/s, respectively. Armstrong et al. (2002) considered that the mass of material ejected from the Earth is about 0.14 of the mass of the impactor at its velocity equal to 14 km/s. Here and below in the paper, if is not mentioned specially, similar values of mass are considered as fractions, not in percent. Based on Figs. 3 and 5 from (Shuvalov and Trubetskaya, 2011), we can conclude that at an impact velocity $v_{imp} = 18$ km/s, the ratio of the mass of the ejected material (with velocities greater than the parabolic velocity on the Earth's surface) to the mass of the impactor was about 0.002 and 0.2 for vertical and oblique (at 45 degrees) impacts, respectively. In these calculations, the impactor and the target were granite. Shoemaker (1962) showed that maximum ejection was at impact angles equal to 45°. Artemieva and Ivanov (2004) studied collisions of bodies with Mars and obtained that ejection was maximum at an impact angle equal to 30°, and was smaller by one or two orders of magnitude at vertical collisions. An ejection angle was smaller for a greater ejection velocity (see fig. 9 in their paper). Svetsov (2011) considered that the ratio of the mass of ejected material to the mass of an impactor with a velocity $v_{imp}$ is proportional to $((v_{imp}/v_{ej})^2 - 1)$. From Fig. 1 from (Svetsov et al., 2012) one can conclude that the mass of material ejected from the Earth was maximum at collision angles of about 30–45 degrees and did not exceed 0.04, 0.06, and 0.13 of the impactor mass at collision velocities of 12, 15, and 20 km/s, respectively. The results of the above papers testify that the mass of the material ejected from the Earth at the considered impacts (not giant impacts) does not exceed 0.2 of the mass of the impactor at $v_{imp} \leq 18$ km/s.

Artemieva and Shuvalov (2008) simulated high-velocity impact ejecta following falls of comets and asteroids onto the Moon. They noted that the total mass of ejected material on average was greater by a factor of 2 than the impactor mass, and the Moon lost about 0.0001 of its mass during the last 3.8–3.9 Gyr. For the Earth, the fraction of ejected mass was smaller and it should not exceed $0.0001m_E$ for this time interval. The Earth lost its material mainly during more older stages of its accumulation.

Shuvalov and Trubetskaya (2011) showed that the values of an ejection angle $i_{ej}$ are mainly between 20° and 55°, with a peak between 40° and 50°. Raduncan et al. (2022) obtained approximately the same values of $i_{ej}$ for collisions of bodies with asteroids. In experiments with micron particles, Barnouin et al. (2019) obtained ejection angles from 40° to 80° at collision velocities $v_{imp} = 1$ km/s and from 43° to 59° at higher $v_{imp} = 2.5$ km/s. Beech et al. (2019) noted that due to the atmosphere, only bodies with a diameter of at least 0.7 m can leave the

Earth. Artemieva (2014) showed that porosity can reduce the ejection of solid fragments, which are the source of lunar meteorites, by an order of magnitude.

Results presented in Section 1.2 show that on average the bodies that came to the Earth from more distant distances from its orbit had greater collision velocities, and characteristic ejection velocities are greater for greater collision velocities. The amount of material ejected from the Earth as a result of a collision of a body with the Earth (exclusive for giant impacts) usually is smaller by at least a factor of several than the mass of an impactor. Ejection angles have a peak at about 45°.

### 1.3. Motion of bodies ejected from the terrestrial planets

Some bodies that were ejected from collisions of impactors with the Earth later could collide with planets, the Sun, or the Moon, or could be ejected from the Solar System. After the Late Heavy Bombardment, falls of large bodies capable of causing the ejection of material from the Earth beyond its Hill sphere are rare. The Hill sphere is the region around the Earth where its own gravity (compared to that of the Sun) is the dominant force in attracting satellites, and it equals to 230.7 Earth radii. Melosh and Tonks (1993) studied the ejection and exchange of surface material among the terrestrial planets. They used Opik-type orbital evolution models and considered ejection from each planet in random directions with velocities at a great distance from the planet of 0.5, 2.5, and 5.0 km/s. It was obtained that Earth ejecta was mainly reaccreted by the Earth, but about 30 % strike Venus within 15 Myr and 5 % strike Mars within 150 Myr after ejection. About 20 % of Earth ejecta was thrown out of the Solar System. Armstrong et al. (2002) proposed to study lunar rocks in order to find material ejected from the Earth during the Late Heavy Bombardment and to better understand the history of the Earth. They discussed the ejecta transfer from the Earth, Venus, and Mars to the Moon. Gattacceca et al. (2023) supposed that the found meteorite Northwest Africa 13,188 is a terrestrial meteorite that fell back to the Earth.

In (Gladman et al., 2005; Reyes-Ruiz et al., 2012) the motion of bodies ejected from the Earth at collisions of impactors with the Earth was studied during time interval equal to 30 kyr after ejection. The initial velocities were taken to be perpendicular to the surface of the Earth. Initial velocities $v_{ej}$ of ejected bodies varied from 11.22 to 16.4 km/s. Initial positions of bodies were distributed over the Earth surface. Bodies started from the height $h = 100$ km above the Earth surface. The probabilities of collisions of bodies with the Earth, Venus, Mars, and the Sun and the fraction of escape bodies were calculated in both these papers. Gladman et al. (2005) noted that for such calculations the swift-rmvs3 symplectic integration code (Levison and Duncan, 1994) gave similar results as other methods of integration which they used, though other methods can take into account close encounters more accurately. In (Gladman et al., 2005), the probability of a collision of an ejected body with the Earth during 30 kyr varied from 0.09 at $v_{ej} = 11.22$ km/s to 0.001–0.003 at $13.2 \leq v_{ej} \leq 16.4$ km/s. Reyes-Ruiz et al. (2012) additionally calculated collisions with the Moon, Jupiter, and Saturn. In their calculations they took into account the gravitational influence of the Sun, all planets, and the Moon. For integration they used the integration package Mercury (Chambers, 1999). In (Reyes-Ruiz et al., 2012), integration using the symplectic algorithm (with a step of 24 h) was replaced by integration using the Bulirsh-Stern integrator (with an accuracy at a step equal to $10^{-10}$) when the bodies approached massive objects. Reyes-Ruiz et al. (2012) studied the effect of ejection location on the probabilities of collisions of bodies with planets. They calculated such probabilities during 30 kyr for ejection velocities equaled to 11.22, 11.71, 12.7, 14.7, and 16.4 km/s. For example, at $v_{ej} = 12.7$ km/s the fractions of bodies collided with the Earth, Moon, Venus, and Mars equaled to 0.0034, 0.0001, 0.00085, and 0.00003, respectively. For comparison, for a smaller number of considered ejected bodies Gladman et al. (2005) obtained these fractions for the Earth and Venus to be equal to 0.004 and 0.001. Collisions of ejected bodies with Mercury were not calculated in (Gladman et al., 2005; Reyes-Ruiz et al., 2012). In these papers, the considered time interval equaled to 30 kyr because for this time biological material can remain viable in space, may return to the Earth and "reseed" terrestrial life after the sterilizing effect of a giant impact has passed. Armstrong et al. (2002) noted that most of bodies ejected from the Earth at a near-parabolic velocity fell onto the Earth or the Moon and most of such collisions were during 100 years.

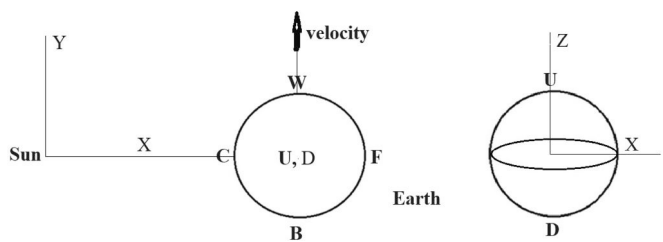

**Fig. 1.** Points of ejection on the Earth's surface. The Earth moves in the plane XY, and Z axis is perpendicular to this plane. Points F and C are on the line from the Sun to the Earth at most and least distant points of the Earth's surface from the Sun, respectively. Point W is the forward point in the direction of the Earth's motion, and point B is on the opposite (to the motion) side of the Earth. Points U and D are located on the Z axis from the center of the Earth with the maximum and minimum values of $Z$, respectively.

The motion of bodies ejected from Mars, Mercury, and the Moon was also studied. Wetherill (1984) considered the evolution of the orbits of bodies ejected from Mars in random directions. The velocity of ejection of bodies from the surface of Mars varied from 5 to 7 km/s. The probabilities of collisions of bodies with the Earth were calculated using Öpik's formulas. Using the symplectic integration method, Gladman et al. (1995, 1996a, 1996b) and Gladman (1997) simulated the evolution of the orbits of bodies ejected from random positions on the surface of the Moon (with ejection velocities of 2.3 and 3.5 km/s), Mars (with the values $v_{\infty} = (v_0^2 - v_{esc}^2)^{1/2}$ of a velocity at the infinity from 1 to 3.3 km/s, where $v_{esc}$ is the escape (parabolic) velocity) and Mercury (with $v_{\infty} = 1$ km/s) and studied characteristic times elapsed between ejections and collisions of these bodies with the Earth and the fraction of bodies that fell onto the Earth. 2100 initial ejected bodies were considered for ejection from Mars, and 200 bodies - for ejection from Mercury. Considered time intervals reached 10, 30, and 50 Myr for the Moon, Mercury, and Mars respectively. By numerically integrating the equations of motion, Gladman and Coffey (2009) studied the planetary-influenced motion of bodies ejected from Mercury toward the Earth and Venus. The velocities of bodies ejected radially from Mercury were taken to be 4, 9, 14, 20, and 25 km/s. It was found that 2–5 % of the ejected substance reached the Earth in 30 million years.

Emsenhuber and Asphaug (2019) and Emsenhuber et al. (2021) studied the motion of bodies ejected as a result of giant impacts with the Earth and Venus. The masses of impactors were equal to $0.15m_E$ in (Emsenhuber et al., 2021) and were even larger ($0.2m_E$ and $0.5m_E$) in (Emsenhuber and Asphaug, 2019). Emsenhuber et al. (2021) noted that there were no ejected bodies that traveled from the Earth to Venus or from Venus to the Earth before 0.1 Myr. In their Table 3 the fractions of bodies ejected from the Earth that later collided with the Earth and Venus equaled to 0.47–0.68 (the range is for different impact angles) and 0.16–0.3 at $v_{imp}/v_{esc} = 1.1$ and to 0.3–0.4 and 0.31–0.32 at $v_{imp}/v_{esc} = 1.2$, where $v_{imp}$ is the impact velocity. For bodies ejected from Venus, the above ratios were equal to 0.13–0.17 and 0.7–0.78 at $v_{imp}/v_{esc} = 1.1$ and to 0.11–0.31 and 0.5–0.8 $v_{imp}/v_{esc} = 1.2$ (see Table 4 in (Emsenhuber et al., 2021)). There was an exchange of ejected material between the Earth and Venus.

Mileikowsky et al. (2000) and Bradak (2023) discussed putative microbial exchange between early Earth and Mars. Bradak (2023) concluded that for a 20 km diameter size impactor the average ejectile size is around 1 m diameter at 30 km/s impact velocity. Carter and Stewart (2022) concluded that a portion of late-accreted mass came to the Earth from fragments of intermediate planetary embryos ejected earlier in the history of the Solar System, rather than from primitive chondritic planetesimals.

### 1.4. Problems considered in the paper

Below in the paper, the motion of bodies ejected from the Earth is studied under the gravitational influence of the Sun and all present planets. This motion was simulated just after the formation of the terrestrial planets (with the present orbits and masses of all planets) until all bodies collided with the Sun or planets or reached 2000 AU from the Sun. The considered time interval for some bodies could exceed 1000 Myr and in my calculation variants with 250 ejected bodies typically was about a few hundred Myr and was by four orders of magnitude larger than 30 kyr considered in (Gladman et al., 2005; Reyes-Ruiz et al., 2012). The above authors studied the motion of ejected bodies started vertically from the height of 100 km above the Earth's surface. We considered different ejection angles for bodies started from the surface of the Earth. Initial data and algorithms used for calculations are discussed in the Section 2. The main aim of the paper was to study the probabilities of collisions of bodies ejected from the Earth with planets. These probabilities are discussed in the Section 3. The amount of material ejected from the Earth that could be delivered to planets is discussed in Section 4.

## 2. Initial data and algorithms used for calculations

In each calculation variant, the motion of 250 bodies ejected from the Earth was studied for the fixed values of an ejection angle $i_{ej}$ (measured from the surface plane), a velocity $v_{ej}$ of ejection, a point of ejection, and a time step $t_s$ of integration. The velocity of rotation of the Earth's surface (0.46 km/s at the equator) was not taken into account, as it is much smaller than an ejection velocity. The gravitational influence of the Sun and all eight present planets was taken into account. The calculations ended at time $T_{end}$ when all bodies had collided with planets or the Sun or had reached 2000 AU from the Sun (no bodies left in simulations). In most variants, the values of $T_{end}$ equaled a few hundred Myr (see Section 3.2). Note that bodies were excluded from integration when they reached 30 AU in (Gladman et al., 2005) and 40 AU in (Reyes-Ruiz et al., 2012). The probabilities of collisions of ejected bodies with planets were calculated as the number of bodies colliding with the considered planet to the initial number of bodies.

The ejection of bodies from six opposite points (see Fig. 1) of the Earth's surface was considered for a number of values of velocities and angles of ejection of bodies. In the $vf$ and $vc$ series $v_s$ of calculations (for the starting points F and C), the motion of the bodies started at most and least distant points of the Earth's surface from the Sun (located on the line from the Sun to the Earth), respectively. In the $vw$ and $vb$ series (for the points W and B), the bodies started from the forward point on the Earth's surface in the direction of the Earth's motion and from the back point on the opposite side of the Earth, respectively (from apex and antapex). In the $vu$ and $vd$ series (for the points U and D), the bodies started from points on the Earth's surface with the maximum and minimum values of $z$ (with the z axis perpendicular to the plane of the Earth's orbit), respectively.

Impactors can fall onto different regions of the Earth's surface with different frequency, and so probabilities of ejection from different points on the Earth can be different. Gallant et al. (2009) showed that the ratio of the number of impactors within $30^o$ of the pole to that of a $30^o$ band centered on the equator is about 0.958 for the Earth and 0.903 for the Moon. Zuluaga and Sucerquia (2018) studied impact positions of bodies colliding with the Earth. They concluded that the lowest fluxes of bodies falling onto the Earth are from apex (the direction of the Earth's motion) and antapex, while the largest number of incoming bodies are from perpendicular directions. Location at $60-90^o$ from the apex are more prone to impacts. Numerical simulations by Gallant et al. (2009) and Ito and Malhotra (2010) showed that a leading/trailing hemispherical ratio for lunar impacts by near-Earth objects equals to 1.3, and the average impact velocity of projectiles tends to be larger on the leading side than on the trailing side. A leading/trailing asymmetry of lunar craters was also discussed in (Wang, 2016).

In my different variants, the values of the ejection angle $i_{ej}$ were $15^o$, $30^o$, $45^o$, $60^o$, $89^o$, or $90^o$. The choice of $i_{ej} = 89^o$ was caused by that vectors of initial velocities were very close to each other at $i_{ej} = 90^o$ in each calculation variant (see Section 3.3). Note that in (Gladman et al., 2005; Reyes-Ruiz et al., 2012) considered initial velocities were perpendicular to the surface of the Earth (i.e., $i_{ej} = 90^o$). In my calculations, the ejection velocity $v_{ej}$ was mainly equal to 11.22, 11.25, 11.3, 11.4, 11.5, 12, 14, 16.4, or 20 km/s. In some series of calculations, also other values of $v_{ej}$ between 11.25 and 11.3 km/s were considered in order to study the dependence of the probabilities of collisions of ejected bodies with the Earth on $v_{ej}$ for this interval of values of $v_{ej}$, as these probabilities can change by a factor of 2 for such $v_{ej}$. Different parameters for initial data are summarized in Table 1.

At the fixed values of $i_{ej}$ and $v_{ej}$, 250 different possible orientations of the vector of a velocity of ejection were considered. These vectors belonged to the cone which axis is perpendicular to the Earth' surface, and projections of the vectors onto the surface were on rays from the point of ejection. The directions of two close rays differed by the angle equal to $360/250 = 1.44$ degree. The total number of considered ejected bodies was greater than 75,000. For a few of them, calculations were made with different integration time steps $t_s$ (see Section 3.1) in order to understand whether the obtained results and conclusions can depend on $t_s$. In most calculations, bodies started directly from the surface of the Earth. If it is not mentioned specially, the obtained results are presented for such calculations. Some calculations were made at a few values of the height $h$ of the point of ejection above the Earth in order to study the dependence of results on $h$, as calculations by Gladman et al. (2005) and Reyes-Ruiz et al. (2012) were made for $h = 100$ km, and most of my calculations are for $h = 0$. The results of calculations made for different values of $h$ are discussed in Section 3.8. In the considered model, the ejected material was in the form of bodies. For giant impacts, a disk of material was ejected in the form of gas and dust and formed a disk around the Earth (Canup, 2012; Cuk and Stewart, 2012; Rufu et al., 2017). Therefore, the results of calculations of ejection of bodies presented in my paper do not correspond to the ejection of gas and dust at such giant impacts.

My calculations of the motion of bodies ejected from the Earth and of the probabilities of their collisions with planets were made for the present orbits and masses of planets. Such masses and orbits of the terrestrial planets could differ a little from the values at the late stages of formation of the terrestrial planets. The obtained probabilities of collisions of ejected bodies with the terrestrial planets probably could give similar values as for the case when Jupiter and Saturn got almost present orbits and masses, but Uranus and Neptune could still increase their masses due to accumulation of planetesimals. I hope that such probabilities do not differ much from the values at the late stages of gas-free accumulation of the terrestrial planets and at the Late Heavy Bombardment even at not large time variations of masses and orbits of planets. For the conclusions made below in the paper, it is not needed to calculate accurately the probabilities of collisions of bodies with planets.

The rmvs-3 symplectic code from the SWIFT integration package (Levison and Duncan, 1994) was used for integration of the motion equations. Frantseva et al. (2022) noted that for this code, a time step of integration is reducing considerably at a distance less than 3.5 Hill radii. The calculation results presented in (Ipatov and Mather, 2004a, 2004b) showed that the symplectic algorithm and the BULSTO algorithm (Bulirsh and Stoer, 1996) gave similar results when studying the motion of bodies in the Solar System. Nesvorný et al. (2021a) calculated the motion of bodies under the gravitational influence of planets except Mercury with the use of the rmvs code and obtained that the main results of calculations were about the same for calculations with a 1-day timestep and with a 3-day timestep.

## 3. Results of calculations

### 3.1. Results of calculations with different values of an integration time step

The results of my calculations with different integration time steps $t_s$

**Table 1**
The values of velocities $v_{ej}$ of ejection, angles $i_{ej}$ of ejection, and points $N_{ej}$ of ejection for which calculations have been made for all their possible combinations ($v_{ej}$, $i_{ej}$, $N_{ej}$).

| points $N_{ej}$ of ejection | B, C, D, F, U, W |
|---|---|
| $v_{ej}$ (in km/s) | 11.22, 11.25, 11.3, 11.4, 11.5, 12, 14, 16.4, 20 |
| $i_{ej}$ (in degrees) | 15, 30, 45, 60, 89, 90 |

(equal to 1, 2, 5, or 10 days) were compared and showed possibilities of collisions of bodies with planets at $t_s \le 5^d$. Most of calculations were made with an integration time step $t_s$ equal to 5 days. If it is not noted specially, the results presented in the paper were obtained at $t_s = 5^d$. The probabilities of collisions of ejected bodies with the Earth ($p_e$), Venus ($p_v$), Mercury ($p_{me}$), Mars ($p_{ma}$), and the Sun ($p_{sun}$), and the probability $p_{ej}$ of ejection of a body into a hyperbolic orbit as functions of $i_{ej}$ are presented in Fig. 2 for a whole considered time interval $T_{end}$. Similar plots for the probabilities at a time interval $T = 10$ Myr can be found in Fig. S1 in the Supplementary material. In this paper obtained probabilities are considered as fractions (not percent). The values of the probabilities are presented in these figures as functions of $i_{ej} + i_{ts}$, where $i_{ts}$ is equal to -4°, −2°, 0 or 2° for an integration time step $t_s$ equal to $1^d$, $2^d$, $5^d$, or $10^d$, respectively. The use of $i_{ts}$ is needed for presentation of the probabilities at an integration time step $t_s$ not equal to $5^d$. The probabilities presented at different values of $i_{ts}$ correspond to the same angle of ejection $i_{ej}$, but for different integration time steps $t_s$. Different plots are for different ejection velocities $v_{ej}$. Different signs in the figures correspond to the ejection points B, C, D, F, U, and W (to the series *vb*, *vc*, *vd*, *vf*, *vu*, and *vw*). For $i_{ej} = 45°$ and $t_s = 1^d$, the probabilities $p_e$, $p_v$, $p_{me}$, $p_{ma}$, $p_{sun}$, and $p_{ej}$ are presented in Fig. 3 as functions of time, which varied from 0.01 Myr to $T_{end}$. The designations for this figure are the same, as for Fig. 2. For studies of the probabilities of collisions of bodies with planets it is important to know not only their final values, but also how these probabilities varied with time. Consideration of Figs. 3 and S1 helps to understand variations of the probabilities with time.

Based on the figures it is possible to conclude that the values of the probability $p_e$ of a collision of a body with the Earth don't have a tendency to increase or decrease at smaller values of $t_s$, and the values of $p_e$ do not differ much at different $t_s \le 10^d$. Similar conclusions can be made for other probabilities ($p_v$, $p_{me}$, $p_{ma}$, $p_{sun}$, $p_{ej}$) at $t_s \le 5^d$. Therefore, for our studies we can use results of calculations with $t_s = 5^d$.

There can be differences for some calculations with $t_s = 10^d$ from calculations at $t_s \le 5^d$. They are presented below. For the entire considered time interval for the series *vb* at $v_{ej}$ equal to 12 or 14 km/s, the probabilities $p_v$ of collisions of bodies with Venus were a little smaller at smaller $t_s$ than at $t_s = 10^d$. The values of $p_{me}$ and $p_v$ at $t_s = 10^d$ were greater than those at $t_s \le 5^d$ for ejection of bodies from the point B with $v_{ej} = 16.4$ km/s. Greater values of the probability $p_{me}$ of a collision of a body with Mercury at $t_s = 10^d$ compared to those at smaller $t_s$ were also at $v_{ej} = 11.5$ and $v_{ej} = 12$ km/s for the points B and F at $T = T_{end}$ (see Fig. 2c). The differences in the values of $p_{me}$ obtained at different values of $t_s$ typically did not exceed 0.02. The probabilities $p_{sun}$ of collisions of bodies with the Sun two points of ejection and for $v_{ej}$ equal to 12 or 14 km/s were a little smaller at $t_s = 10^d$ than at $t_s \le 5^d$ (see Fig. 2e).

The evolution of orbits of bodies is chaotic because of close encounters of bodies with planets. Therefore, the considered probabilities can differ for calculations with the same $t_s$ but with slightly different initial data. For $v_{ej} = 16.4$ km/s, $T = T_{end}$, and the point F, the values of $p_e$ at $t_s = 2^d$ were smaller than at $t_s$ equal to $1^d$, $5^d$, or $10^d$. In our studies, we do not need to know very accurately the orbital evolution of bodies at their close encounters with planets. The main aim of our calculations was to calculate the probabilities of collisions of bodies with planets. As these probabilities are about the same at $1^d \le t_s \le 5^d$, then there was no need to use to more accurate (but slower) methods of integration.

### 3.2. Times of motion of ejected bodies

In my calculation variants, the motion of bodies ejected from the Earth was studied during the maximum dynamical lifetime $T_{end}$ for all bodies, which was mainly about 200–500 Myr. For a few variants it exceeded 1000 Myr. These values of $T_{end}$ are greater by about 4 orders of magnitude than 30 kyr used by Gladman et al. (2005) and Reyes-Ruiz et al. (2012). In different calculation variants, the last body can collide with any terrestrial planet. Variants with $T_{end} < 100$ Myr include the calculations with $v_{ej} \le 11.25$ km/s (when all bodies quickly fell back

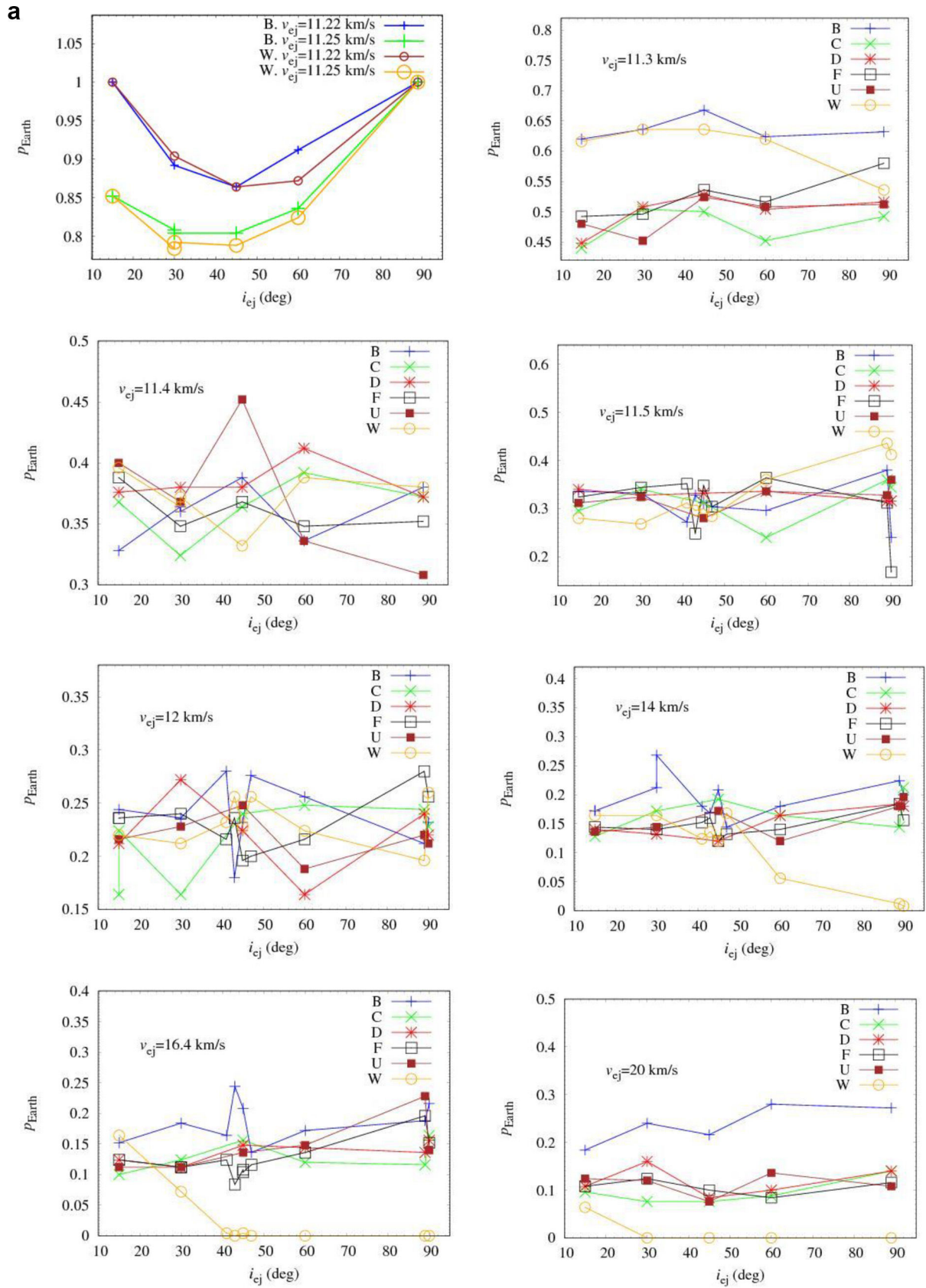

**Fig. 2.** The fractions of bodies that collided with the Earth ($p_e$, Fig. 2a), Venus ($p_v$, Fig. 2b), Mercury ($p_{me}$, Fig. 2c), Mars ($p_{ma}$, Fig. 2d), the Sun ($p_{sun}$, Fig. 2e) or ejected into hyperbolic orbits ($p_{ej}$, Fig. 2f) during the entire considered time interval vs. the ejection angle $i_{ej}$ at several values of an ejection velocity $v_{ej}$ and six points of ejection (B, C, D, F, U, and W). The values of the fractions are presented for $i_{ej} + i_{ts}$, where $i_{ts}$ is equal to -4°, −2°, 0, or 2° for $t_s$ equal to $1^d$, $2^d$, $5^d$, or $10^d$, respectively. Each sign on the figure corresponds to the mean value for 250 bodies.

**b**

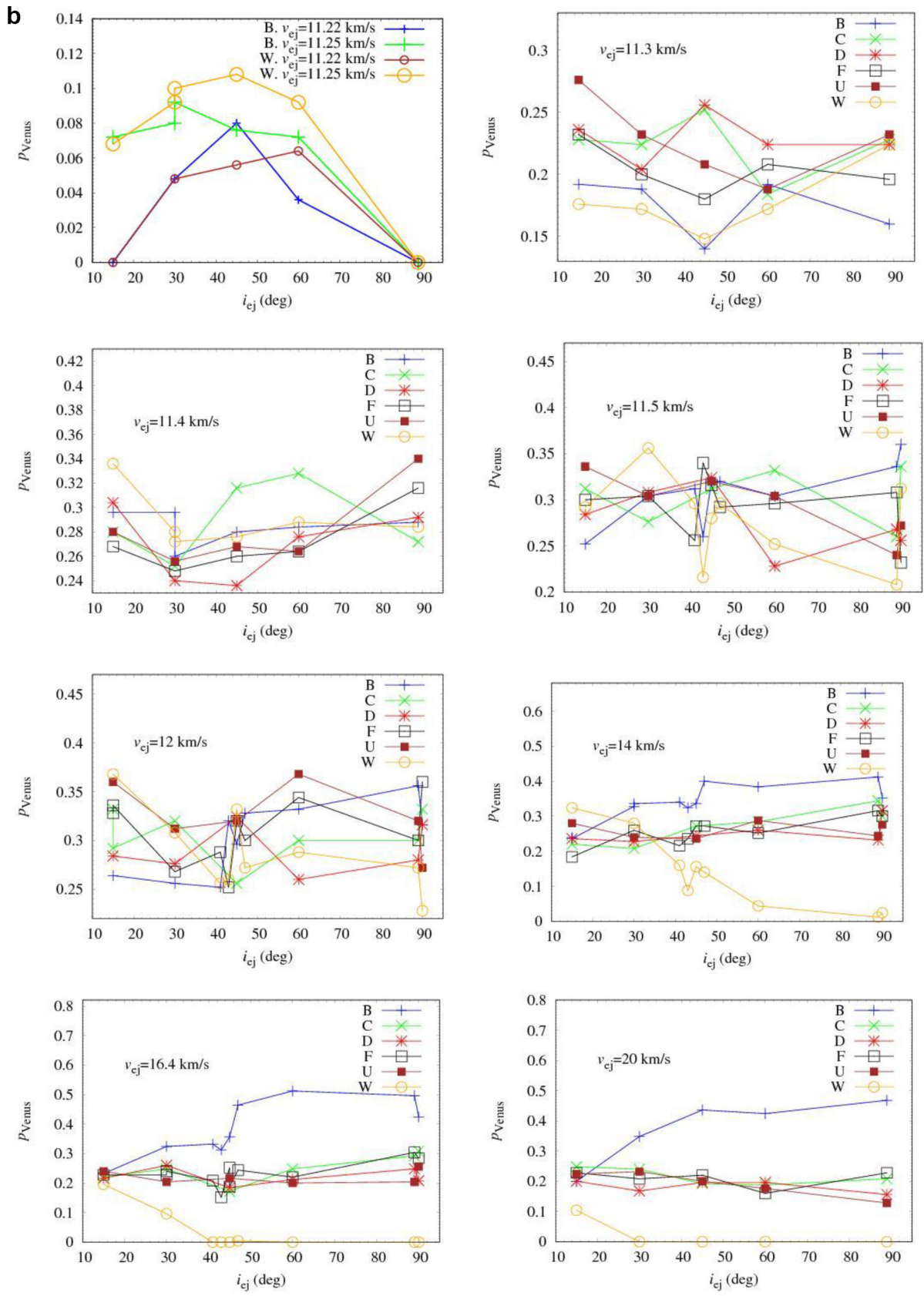

**Fig. 2.** (*continued*).

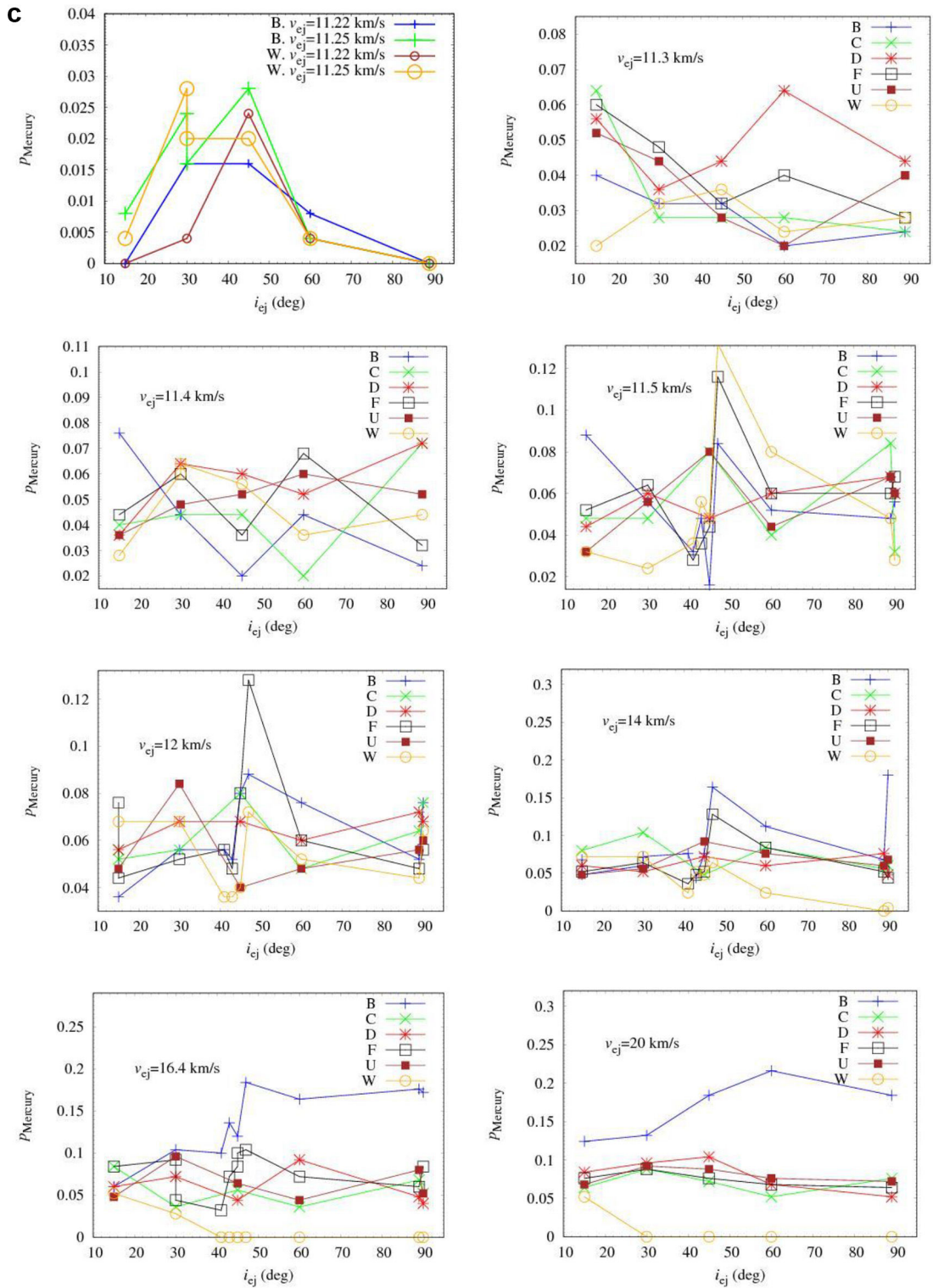

Fig. 2. (*continued*).

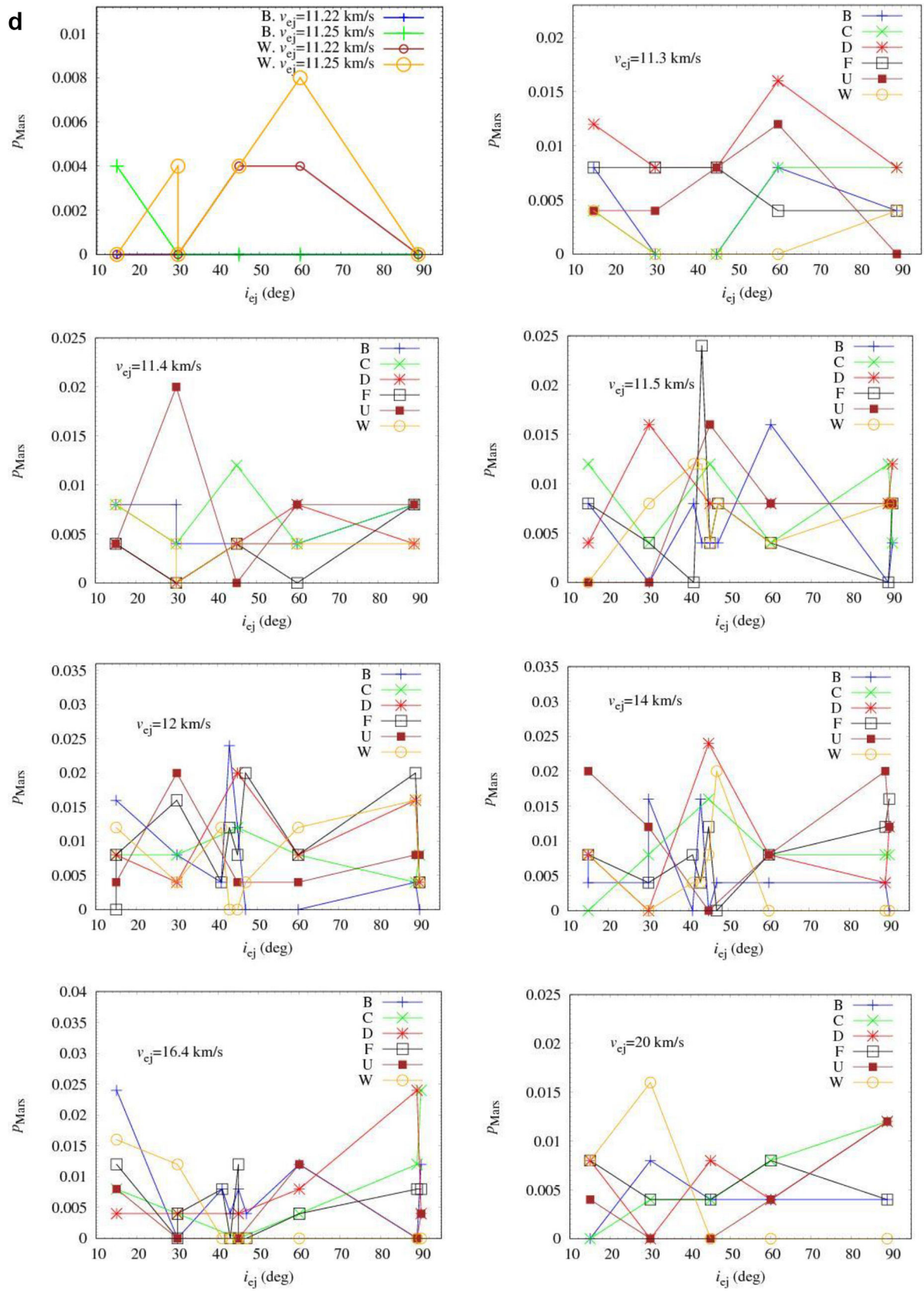

**Fig. 2.** (*continued*).

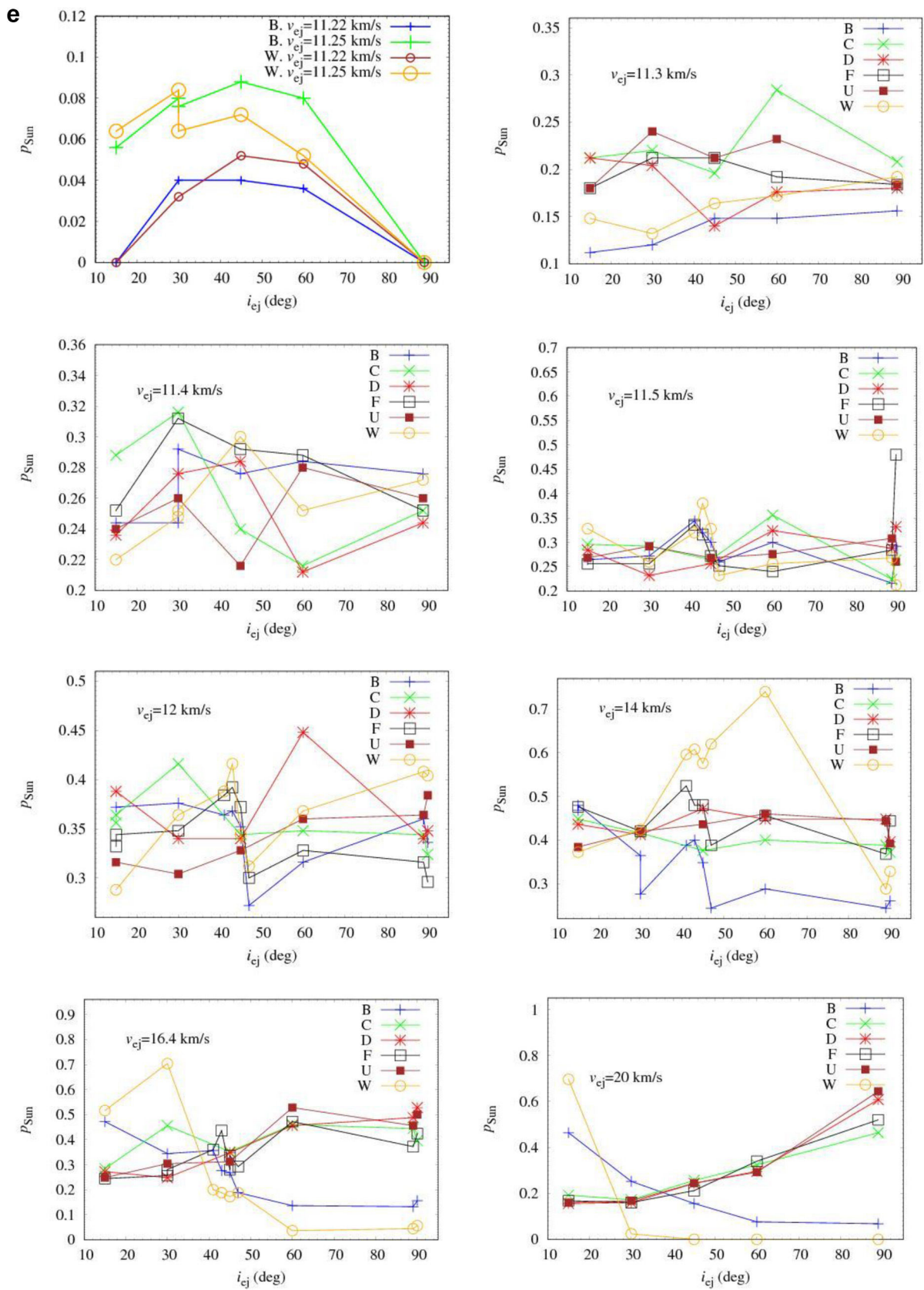

**Fig. 2.** (*continued*).

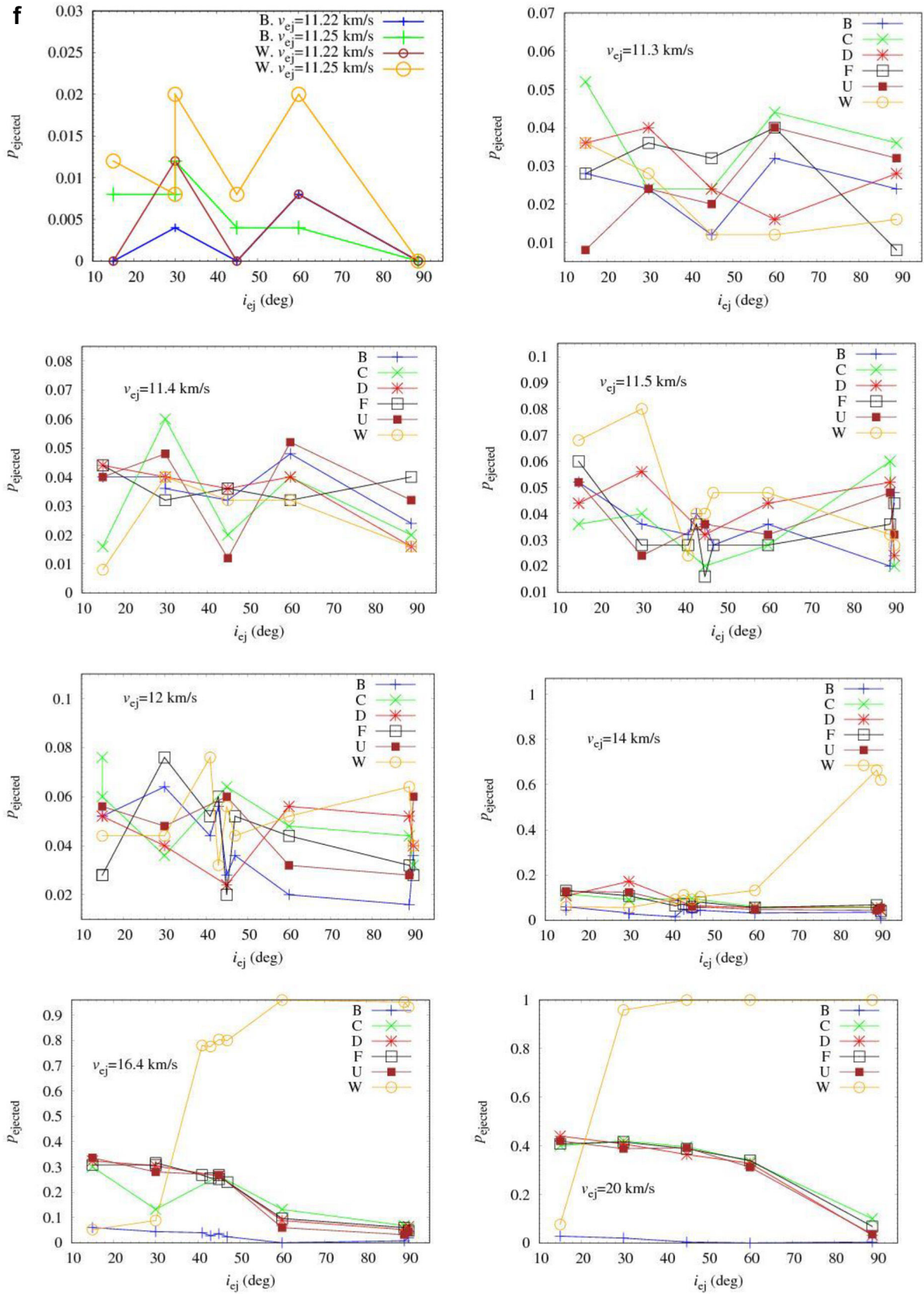

**Fig. 2.** (*continued*).

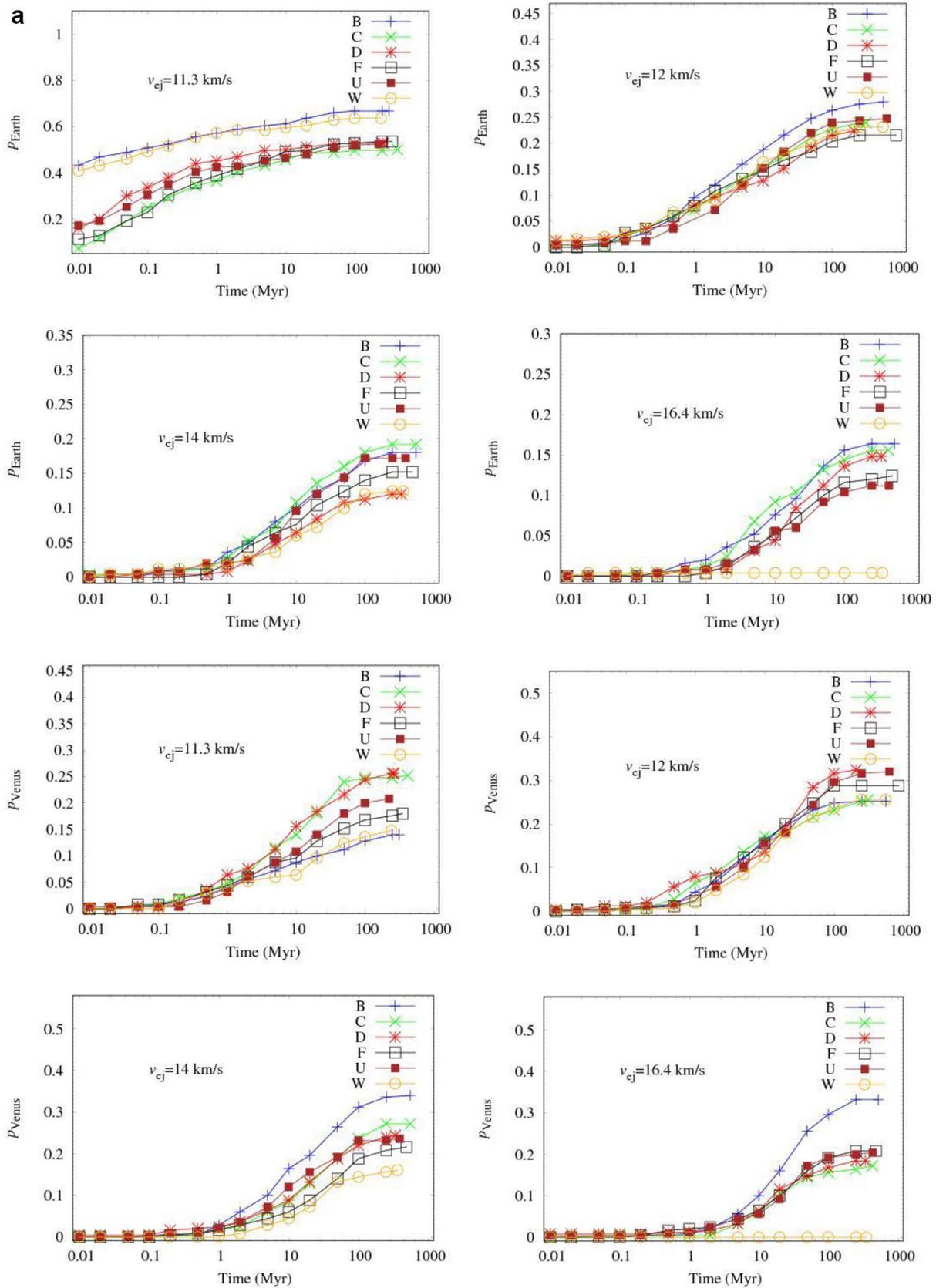

**Fig. 3.** The fractions of bodies that collided with the Earth ($p_e$, Fig. 3a), Venus ($p_v$, Fig. 3a), Mercury ($p_{me}$, Fig. 3b), Mars ($p_{ma}$, Fig. 3b), the Sun ($p_{sun}$, Fig. 3c) or ejected into hyperbolic orbits ($p_{ej}$, Fig. 3c) as functions of time, which varied from 0.01 Myr to $T_{end}$, for $i_{ej}$ = 45°. Different plots are for different ejection velocities $v_{ej}$. Different signs and lines in the figures correspond to the ejection points B, C, D, F, U, and W. Each sign on the figure corresponds to the mean value for 250 bodies.

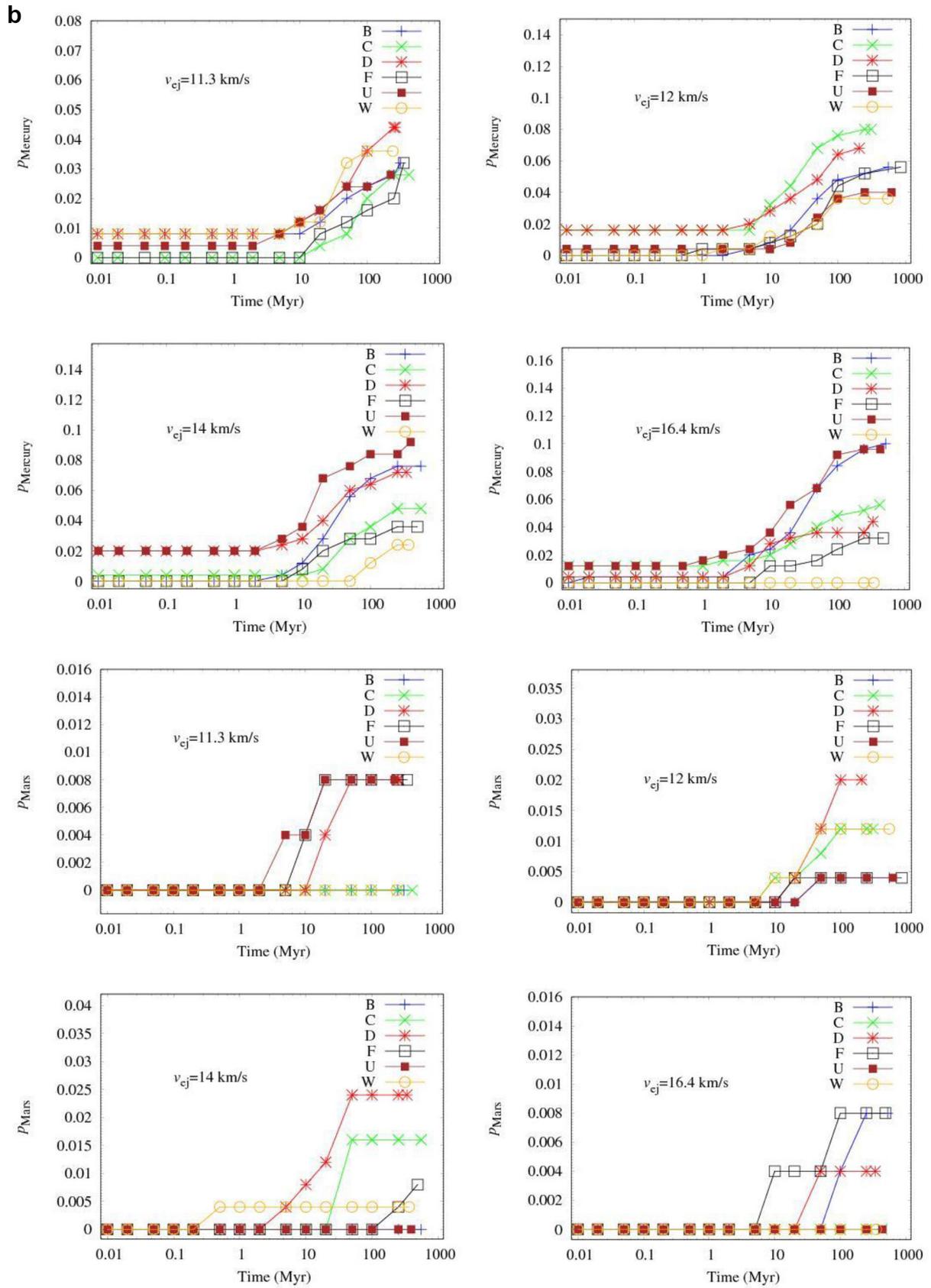

**Fig. 3.** (*continued*).

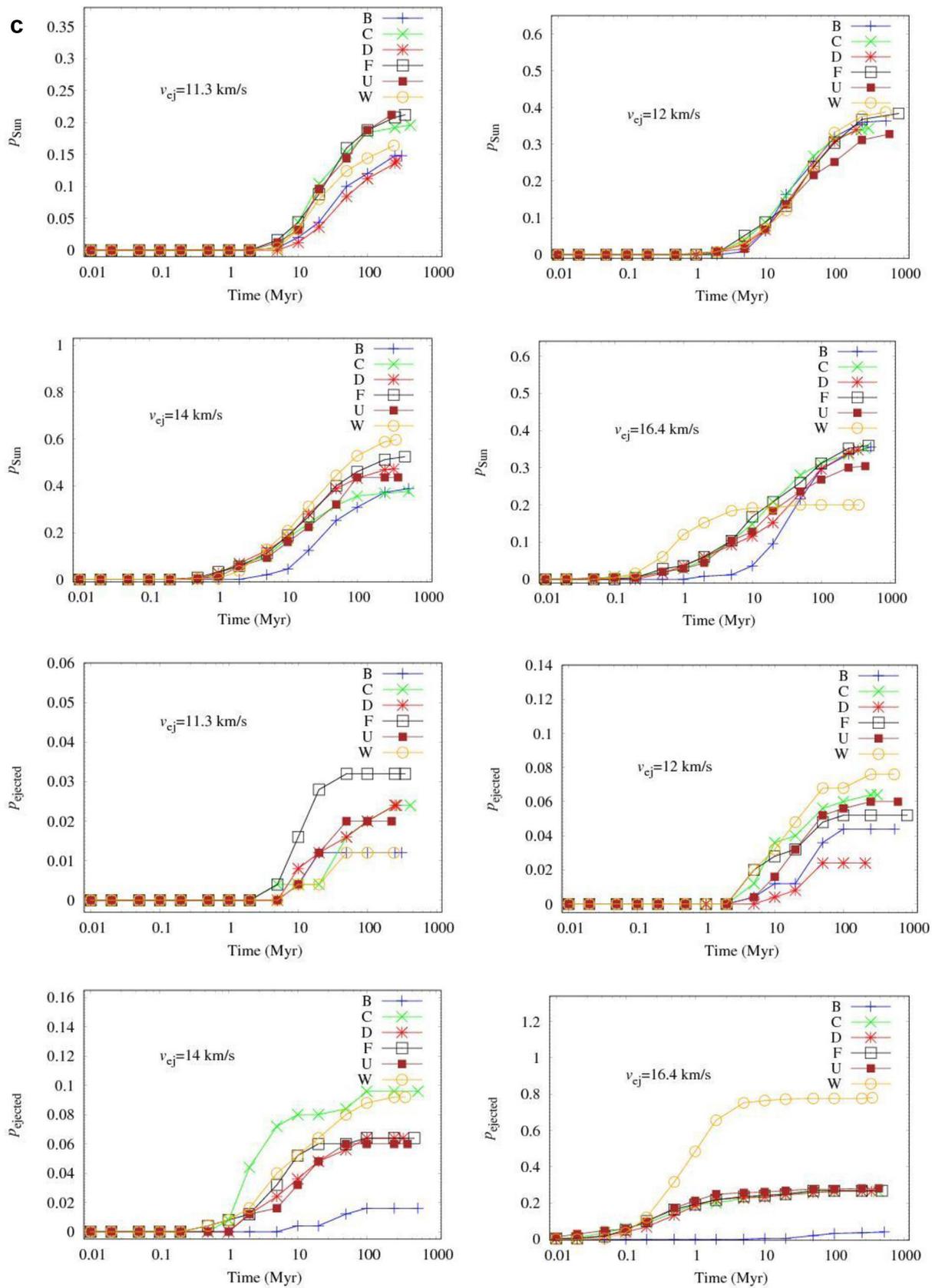

**Fig. 3.** (*continued*).

**Table 2**

The fraction $p_{el}$ of bodies still moved in elliptical orbits at time $T$ (in Myr) for several ejection velocities $v_{ej}$ (in km/s) at an ejection angle equal to 45°. The range of the values of $p_{el}$ is for different starting points.

| $T$ (Myr) \|\| $v_{ej}$ (km/s) | 11.3 | 11.5 | 12 | 14 | 16.4 |
|---|---|---|---|---|---|
| 0.1 | 0.65–0.75 | 0.90–0.96 | 0.97–0.99 | 0.97–1.0 | 0.92–1.0 |
| 1 | 0.39–0.55 | 0.74–0.79 | 0.86–0.90 | 0.91–0.99 | 0.76–0.95 |
| 10 | 0.32–0.38 | 0.46–0.57 | 0.55–0.61 | 0.55–0.64 | 0.48–0.74 |
| 100 | 0.03–0.06 | 0.04–0.1 | 0.07–0.12 | 0.09–0.15 | 0.07–0.1 |

onto the Earth) and also the ejection of bodies from the point W with $v_{ej} \geq 16.4$ km/s.

At ejection velocities close to the escape velocity, all bodies could fall back onto the Earth in several tens of days. For example, all bodies fell back onto the Earth during not more than 15–50 days for several variants at $v_{ej} = 11.22$ km/s (e.g., for the *vb* variants at $i_{ej} = 15°$ or $i_{ej} = 89°$, for the *vc* variants at $i_{ej} = 45°$, for the *vf* variants at $i_{ej} = 15°$ or $i_{ej} = 89°$, and for the *vw* variants at $i_{ej} = 15°$). It was obtained that $T_{end} \leq 50^d$ for some variants at $v_{ej} = 11.25$ km/s (e.g., for the *vb* variants at $i_{ej} = 89°$, for the *vc* variants at $i_{ej} = 60°$, for the *vd* variants at $i_{ej} = 30°$, for the *vf* variants at $i_{ej} = 89°$, and for the *vu* variants at $i_{ej} = 30°$ and $i_{ej} = 89°$). For the starting point D, $T_{end}$ equaled to $165^d$ at $v_{ej} = 11.25$ km/s and $i_{ej} = 45°$. The value of $T_{end}$ equaled to a few years at $v_{ej} = 11.25$ km/s for the *vd* variants at $i_{ej} = 45°$ or $i_{ej} = 60°$, for the *vc* variants at $i_{ej} = 30°$ or $i_{ej} = 45°$, and for the *vu* variants at $i_{ej} = 45°$. These calculations show that small values of $T_{end}$ were more often at $i_{ej} = 15°$ or $i_{ej} = 89°$ then for $i_{ej} = 45°$. The values of $T_{end}$ were between 1 and 12 Myr when all or almost all bodies were ejected into hyperbolic orbits in the variants *vw* at $v_{ej} = 16.4$ km/s and $i_{ej} \geq 60°$ and also at $v_{ej} = 20$ km/s and $i_{ej} \geq 45°$.

The dynamical lifetime of one body at the series $v_w$, $v_{ej} = 14$ km/s, and $i_{ej} = 89°$ equaled to 5005 Myr. This body moved for a long time in the trans-Neptunian region in a Neptune-crossing orbit. In principle, such Neptune-crossing bodies could have a chance to change their orbits due to gravitational influence of other trans-Neptunian bodies and left in the trans-Neptunian belt. Of course, the fraction of trans-Neptunian objects originated in the feeding zone of the terrestrial planets could be very small. At the series $v_u$, $v_{ej} = 11.5$ km/s, and $i_{ej} = 30°$, the dynamical lifetime of the last body equaled to 1146 Myr. Before a collision with the Sun this body moved some time in the inner part of the main asteroid belt. In that variant after $T = 100$ Myr, five bodies collided with the Sun, one body collided with Venus, and one body collided with Mercury.

Note that the results of evolution of disks corresponding to the feeding zone of the terrestrial planets presented in (Ipatov, 1987) showed that for highly inclined orbits the times of last collisions of planetesimals with the Earth could reach 500 Myr. In (Ipatov, 1993; Marov and Ipatov, 2023) it was noted that the time to form 80 % of the Earth from a disk of planetesimals did not exceed 10 Myr, and the Earth and Venus could acquire 50 % of their masses in 5 Myr. In calculations by Nesvorný et al. (2023), the Earth grew to a half of its final mass, and Venus grew to 85 % of its final mass in only 3 Myr.

The fraction $p_{el}$ of bodies that still moved in elliptical orbits (i.e., were not ejected from the Solar System and have not collided with planets or the Sun) at time $T$ (in Myr) for several ejection velocities $v_{ej}$ at an ejection angle $i_{ej}$ equal to 45° is presented in Table 2. The range of the values of the fraction is for different starting points. The values of $p_{el}$ presented in Table 2 are mainly greater for greater ejection velocities. At $v_{ej} \geq 11.5$ km/s, the fraction $p_{el}$ of bodies left in elliptical orbits was more than 0.9, 0.74, and 0.46 at time $T$ equal to 0.1, 1, and 10 Myr, respectively. On average, about a half of bodies still moved in elliptical orbits at $T = 10$ Myr and less than 10 % of initial ejected bodies still moved in elliptical orbits at $T = 100$ Myr. It is possible to see from Fig. 3 that a few bodies collided with planets after 250 Myr. The rightmost signs in Fig. 3 show maximum dynamical lifetimes of bodies in each considered variant. In this figure all such maximum lifetimes are greater than 100 Myr. As it is discussed at the beginning of this Section 3.2, small lifetimes could be only at ejection velocities close to the escape velocity.

### 3.3. Probabilities of the collisions of bodies ejected from the Earth with the Earth

Probabilities of the collisions of bodies ejected from the Earth with the terrestrial planets at $T = T_{end}$ and $T = 10$ Myr are presented in Figs. 2 and S1 for different values of $v_{ej}$ and $i_{ej}$ at different points of ejection. The ranges of such probabilities for different considered $i_{ej}$ and four points of ejections (C, D, F, and U) are presented in Tables 3 and 4 at a few values of $v_{ej}$ for $T = 10$ Myr and $T = T_{end}$, respectively. For the starting points B and W, the values of the probabilities can be outside these intervals.

Probabilities $p_e$ of collisions of the bodies with the Earth were greater for a smaller ejection velocity $v_{ej}$. For example, at $T = T_{end}$ for the points of ejections C, D, F, and U, mean values of $p_e$ varied from about 1 at $v_{ej} \leq 11.25$ km/s to about 0.1 at $v_{ej} = 20$ km/s, with $p_e$ about 0.25–0.35, 0.16–0.27, and 0.1–0.2 at $v_{ej}$ equal to 11.5, 12, and 14 km/s, respectively.

The values of the probabilities $p_e$ can be different for different angles $i_{ej}$ of ejection. At $11.22 \leq v_{ej} \leq 11.25$ km/s for the points B and W, the values of $p_e$ were smaller at $30° \leq i_{ej} \leq 60°$ than at other $i_{ej}$. For these two points at $v_{ej} \geq 14$ km/s, $p_e$ could strongly depend on $i_{ej}$.

Depending on the values of $v_{ej}$ and $i_{ej}$, in some cases probabilities $p_e$ can differ for different points of ejection (e.g. it can be higher for ejection from point B located on the trailing face), but in other calculation variants they can be about the same for different points of ejection. For example, the values of $p_e$ for the points B and W were about the same as for other points at $11.4 \leq v_{ej} \leq 12$ km/s. They could be smaller than at other four points for $11.22 \leq v_{ej} \leq 11.25$ km/s. At $v_{ej} = 11.3$ km/s both in Figs. 2 and S1, the values of $p_e$ for the *vb* and *vw* series were mainly greater by about 0.1–0.15 than for other starting points. For the point W (when bodies were ejected from the forward point of the Earth's motion), the values of $p_e$ equal to 0 at $i_{ej} \geq 45°$ and $v_{ej} \geq 16.4$ km/s. In this case, $\geq 80$ % of the bodies were thrown into hyperbolic orbits, and most of other bodies collided with the Sun. At $v_{ej} = 20$ km/s, for the point B (located on the trailing face) the values of $p_e$ were about 0.2–0.3 and were greater than for other points of ejection. Reyes-Ruiz et al. (2012) also noted that the number of bodies falling back to the Earth is higher for those ejected from the trailing face.

There was not much dependence of $p_e$ on $i_{ej}$ for the points of ejections C, D, F, and U in Figs. 2a and S1a. Exclusions can be for $i_{ej} = 90°$. For $i_{ej}$

**Table 3**

Ranges of values of $p_e$, $p_v$, $p_{me}$, $p_{ma}$, $p_{sun}$, and $p_{ej}$ at $T = 10$ Myr and at several values of ejection velocities $v_{ej}$ (in km/s) for all considered values of an ejection angle and for four (C, D, F, and U) points of ejection.

| $v_{ej}$ (km/s) | $p_e$ | $p_v$ | $p_{me}$ | $p_{ma}$ | $p_{sun}$ | $p_{ej}$ |
|---|---|---|---|---|---|---|
| 11.3 | 0.37–0.52 | 0.08–0.16 | 0–0.03 | 0–0.008 | 0.01–0.06 | 0–0.03 |
| 11.4 | 0.25–0.35 | 0.13–0.2 | 0–0.03 | 0–0.004 | 0–0.03 | 0–0.03 |
| 11.5 | 0.17–0.3 | 0.14–0.22 | 0–0.05 | 0–0.008 | 0–0.1 | 0–0.03 |
| 12 | 0.12–0.2 | 0.13–0.2 | 0–0.03 | 0–0.008 | 0.05–0.1 | 0–0.03 |
| 14 | 0.0–0.1 | 0.05–0.15 | 0–0.04 | 0–0.008 | 0.05–0.25 | 0–0.15 |
| 16.4 | 0.03–0.1 | 0.04–0.14 | 0–0.05 | 0–0.012 | 0.02–0.25 | 0–0.33 |
| 20 | 0.03–0.1 | 0.03–0.12 | 0.02–0.06 | 0–0.004 | 0.02–0.12 | 0–0.4 |

**Table 4**

Ranges of values of $p_e$, $p_v$, $p_{me}$, $p_{ma}$, $p_{sun}$, and $p_{ej}$ at $T = T_{end}$ (typically equal to a few hundred Myr) for the series $vc$, $vd$, $vf$, and $vu$ at several values of ejection velocities $v_{ej}$ (in km/s).

| $v_{ej}$ (km/s) | $p_e$ | $p_v$ | $p_{me}$ | $p_{ma}$ | $p_{sun}$ | $p_{ej}$ |
|---|---|---|---|---|---|---|
| 11.3 | 0.45–0.55 | 0.18–0.25 | 0.02–0.06 | 0–0.016 | 0.1–0.25 | 0.01–0.04 |
| 11.4 | 0.3–0.45 | 0.23–0.33 | 0.02–0.08 | 0–0.02 | 0.2–0.32 | 0.01–0.05 |
| 11.5 | 0.25–0.35 | 0.22–0.33 | 0.02–0.08 | 0–0.016 | 0.22–0.33 | 0.02–0.06 |
| 12 | 0.16–0.27 | 0.25–0.37 | 0.04–0.08 | 0–0.02 | 0.3–0.45 | 0.02–0.08 |
| 14 | 0.1–0.2 | 0.2–0.3 | 0.04–0.1 | 0–0.02 | 0.35–0.5 | 0.03–0.15 |
| 16.4 | 0.1–0.15 | 0.18–0.25 | 0.03–0.1 | 0–0.025 | 0.2–0.5 | 0.05–0.33 |
| 20 | 0.08–0.15 | 0.15–0.25 | 0.05–0.1 | 0–0.012 | 0.2–0.6 | 0–0.4 |

**Table 5**

The fraction $p_e$ of bodies collided with the Earth and the fraction $p_H$ of bodies that had not left the Hill sphere of the Earth at several values of an ejection velocity $v_{ej}$ for the ejection point F, $i_{ej} = 45^o$, and $T = T_{end}$.

| $v_{ej}$, km/s | ≤11.26 | 11.27 | 11.28 | 11.29 | 11.3 | 11.35 | 11.4 | 11.45 |
|---|---|---|---|---|---|---|---|---|
| $p_e$ | 1 | 0.95 | 0.64 | 0.58 | 0.54 | 0.43 | 0.37 | 0.30 |
| $p_H$ | 1 | 0.91 | 0.12 | 0.064 | 0.044 | 0.012 | 0.016 | 0.016 |

**Table 6**

The ratio $f_T$ of the probabilities of collisions of bodies with the Earth or Venus at time $T$ (in Myr) to the probabilities at the end of evolution (at $T = T_{end}$) at several values $v_{ej}$ of an ejection velocity (equal to 11.3, 12, 14, or 16.4 km/s) for bodies started from the points F, B, or W at $i_{ej} = 45^o$. The probabilities $p$ of collisions of bodies with the Earth or Venus at $T = T_{end}$ are presented by the bold lines.

| $T$ (Myr) | point | Earth | Earth | Earth | Earth | Venus | Venus | Venus | Venus |
|---|---|---|---|---|---|---|---|---|---|
| | $v_{ej} =$ (km/s) | 11.3 | 12 | 14 | 16.4 | 11.3 | 12 | 14 | 16.4 |
| 0.01 | F | 0.21 | 0 | 0 | 0 | 0.04 | 0 | 0 | 0 |
| 0.1 | F | 0.43 | 0.13 | 0 | 0 | 0.24 | 0.03 | 0 | 0 |
| 1 | F | 0.72 | 0.37 | 0.13 | 0.03 | 0.53 | 0.08 | 0.07 | 0.10 |
| 10 | F | 0.92 | 0.68 | 0.50 | 0.42 | 0.93 | 0.54 | 0.28 | 0.31 |
| 100 | F | 0.98 | 0.94 | 0.92 | 0.94 | 1.0 | 1.0 | 0.87 | 0.92 |
| $T_{end}$ | F, $p$ | **0.54** | **0.22** | **0.15** | **0.12** | **0.18** | **0.29** | **0.22** | **0.21** |
| 0.01 | B | 0.65 | 0.01 | 0.04 | 0 | 0 | 0 | 0.01 | 0 |
| 0.1 | B | 0.76 | 0.06 | 0.20 | 0 | 0.06 | 0.16 | 0.08 | 0 |
| 1 | B | 0.86 | 0.34 | 0.56 | 0.12 | 0.29 | 0.17 | 0.48 | 0.02 |
| 10 | B | 0.92 | 0.67 | 0.93 | 0.46 | 0.63 | 0.65 | 0.92 | 0.30 |
| 100 | B | 1.0 | 0.94 | 1.0 | 0.95 | 0.91 | 0.98 | 1.0 | 0.89 |
| $T_{end}$ | B, $p$ | **0.67** | **0.28** | **0.18** | **0.16** | **0.14** | **0.25** | **0.34** | **0.33** |
| 0.01 | W | 0.64 | 0.05 | 0 | 0 | 0 | 0 | 0 | 0 |
| 0.1 | W | 0.77 | 0.10 | 0.10 | 1.0 | 0 | 0.02 | 0 | 0 |
| 1 | W | 0.90 | 0.33 | 0.13 | 1.0 | 0.27 | 0.08 | 0 | 0 |
| 10 | W | 0.94 | 0.71 | 0.48 | 1.0 | 0.43 | 0.48 | 0.28 | 0 |
| 100 | W | 1.0 | 0.95 | 0.97 | 1.0 | 0.92 | 0.92 | 0.90 | 0 |
| $T_{end}$ | W, $p$ | **0.64** | **0.23** | **0.12** | **0.004** | **0.15** | **0.26** | **0.16** | **0.0** |

$= 90^o$, vectors of initial velocities were very close to each other, and orbital evolution of some bodies could be close to each other. Therefore, the calculations with $i_{ej} = 89^o$ were more informative than those at $i_{ej} = 90^o$.

At ejection velocities $v_{ej} \leq 11.25$ km/s, i.e., slightly greater than the parabolic velocity, most of the ejected bodies fell back onto the Earth. In the $vf$ variants for $i_{ej} = 45^o$, all bodies fell onto the Earth in less than 10 and 20 years at $v_{ej} = 11.25$ km/s and $v_{ej} = 11.26$ km/s, respectively. The values of $p_e$ were smaller for greater $v_{ej}$. For example, for the $vf$ variants, $i_{ej} = 45^o$, and $T = 1$ Kyr, the probability $p_e$ equaled to 0.9, 0.52, and 0.39 at $v_{ej}$ equal to 11.27, 11.28, and 11.3 km/s, respectively.

At the $vf$ series, $i_{ej} = 45^o$, and $T = T_{end}$, the values of $p_e$ and the fraction $p_H$ of bodies that had not left the Hill sphere (and relatively quickly fell back onto the Earth) are presented in Table 5 for several values of $v_{ej}$ between 11.22 and 11.45 km/s. The values of $p_e$ and $p_H$ are close to 1 only at an ejection velocity close to the escape velocity. The value of $p_H$ did not change much at different $i_{ej}$. For example, for point W, $v_{ej} = 11.25$ km/s, and $T = T_{end}$, $p_H$ equaled to 0.7, 0.62, 0.6, and 0.56 at $i_{ej}$ equal to $15^o$, $30^o$, $45^o$, and $60^o$, respectively. The values of $p_H$ and $p_e$ were equal to 1 at $v_{ej} = 11.25$ km/s and $T = T_{end}$ for other five points of ejection. So, for the leading edge of the Earth (point W) the value of $p_H$ is smaller than for other points of ejection at $v_{ej}$ close to the escape velocity.

The ratio of the values of $p_e$ at $T = T_{end}$ to those at $T = 10$ Myr was usually not more than 2. Similar ratio of $p_e$ for $T = 10$ Myr and $T = 1$ Myr was about 2 at $v_{ej} = 12$ km/s, was less than 2 at $v_{ej} = 11.3$ km/s, and often exceeded 2 at $v_{ej} \geq 14$ km/s. Similar ratio for $T = 1$ Myr and $T = 0.1$ Myr could be more than 2.

At $T = 10$ Myr for all six starting points, the values of $p_e$ were about 0.25–0.35, 0.17–0.3, and 0.12–0.2 at $v_{ej}$ equal to 11.4, 11.5, and 12 km/s, respectively. At the points C, D, F, and U at $T = 10$ Myr, $p_e$ was less than 0.1 at $14 \leq v_{ej} \leq 20$ km/s. At $i_{ej} = 89^o$, $p_e$ was about 0.6 for point W at $v_{ej} = 14$ km/s, and $p_e$ was about 0.2 for point B at $v_{ej} = 20$ km/s.

Time variations of the probabilities $p_e$ of collisions of ejected bodies with the Earth are presented in Fig. 3a for $i_{ej} = 45^o$ and $t_s = 1^d$. Table 6 shows the ratio $f_T$ of the values of $p_e$ at a current time $T$ (from 0.01 Myr to 100 Myr) to the values at $T = T_{end}$ for the starting points F, B, and W at $i_{ej} = 45^o$, and it demonstrates the relative growth of $p_e$. Similar values of $f_T$ are presented for probabilities of collisions of bodies with Venus. At $v_{ej}$ equal to 12, 14, and 16.4 km/s, these values were presented for calculations with an integration time step $t_s = 1^d$ (but the values at $t_s = 5^d$ are about the same). Note that all calculations presented in (Ipatov, 2024a) and most calculations presented in the present paper were made at $t_s = 5^d$. In Section 3.1 it is noted that a step $t_s = 5^d$ is enough for our studies.

The data from Table 6 show that more than 90 % of collisions with the Earth and Venus occurred during the first 100 Myr. For smaller $T$ the values of $f_T$ can be different for different ejection velocities.

As it is seen from Fig. 3a, the probabilities $p_e$ for the points C, D, and U don't differ much from those for the point F. The growth of $p_e$ with time at $T < 1$ Myr was greater for a smaller ejection velocity. For $v_{ej} = 11.3$ km/s already at $T = 10$ Kyr, $p_e$ reached 0.4 for the points B and W and 0.1–0.2 for other four points. The ratio of such values of $p_e$ at $T = 10$ Kyr to their final values was about 0.65 and 0.2, respectively. There were not any collisions of bodies with the Earth before $T = 10$ Kyr for $v_{ej} \geq 14$ km/s at the points F and W. Such results do not contradict to the results presented by Gladman et al. (2005) and Reyes-Ruiz et al. (2012) who studied ejection from a whole surface of the Earth and concluded that the probability of a collision of a body with the Earth did not exceed 0.04 during the first 30 Kyr.

The fraction $p_e$ of bodies collided with the Earth during the first million years at $i_{ej} = 45^o$ for all six started points was about 0.4–0.6, 0.07–0.1, 0.008–0.04, and 0.004–0.02 at $v_{ej}$ equal to 11.3, 12, 14, and 16.4 km/s, respectively. For comparison of these data with data from Table 6, note that lines of Table 6 for different $T$ present the ratio $f_T$ of the probabilities of collisions of bodies with the Earth at time $T$ (in Myr) to the probabilities at the end of evolution (at $T = T_{end}$), because the main aim of Table 6 is to show a relative growth of $p_e$ with time. In order to obtain the values of $p_e$ at times $T$, the values of $f_T$ must be multiplied by the values of $p_e$ at $T = T_{end}$, which are presented by bold lines in Table 6.

The typical ratios of probabilities of collisions of bodies ejected from the Earth with planets at $v_{ej} = 11.5$ km/s to those at $v_{ej} = 16.4$ km/s were about 2, 1.5, 1, 0.8, and 0.7 for the Earth, Venus, Mars, Mercury, and the Sun, respectively. As it is noted at the end of Section 1.2, for the impactors that came from the zone of the giant planets, compared to the accumulation of the Earth from its feeding zone, velocities of their collisions with the Earth (and so typical ejection velocities) were greater. Therefore, for such impactors coming from the zone of the giant planets the ratio $p_v/p_e$ of the probabilities of collisions of bodies ejected from the Earth with Venus to those with the Earth was a little greater and the fraction of bodies ejected into hyperbolic orbits was greater than for the impactors from the feeding zone of the Earth.

### 3.4. Probabilities of collisions of bodies ejected from the Earth with Venus

In Table 4 the probabilities $p_v$ of collisions of bodies ejected from the Earth with Venus **at $T = T_{end}$** were often about 0.2–0.3 and less depended on $v_{ej}$ than $p_e$. For $T = T_{end}$ on average $p_v$ was about $p_e$ at $v_{ej} = 11.5$ km/s, was greater than $p_e$ at $v_{ej} \geq 12$ km/s, and it was less than $p_e$ at $v_{ej} \leq 11.4$ km/s. In general, the ratio $p_v/p_e$ of the probability $p_v$ of a body collision with Venus to its probability $p_e$ of a collision with the Earth was smaller at lower ejection velocities. The ratio of $p_v/p_e$ was typically smaller at smaller $T$. The total number of bodies delivered to the Earth and Venus probably did not differ much.

In the variants *vf, vc, vu,* and *vd* at $T = 10$ Myr, the probabilities $p_v$ of a collision of a body with Venus were in the ranges of 0.14–0.22, 0.13–0.2, and 0.03–0.15 at $v_{ej}$ equal to 11.5, 12, and 14–20 km/s, respectively. For the series *vw* at some values of $v_{ej}$ and $i_{ej}$ (e.g. at $v_{ej} = 20$ km/s and $i_{ej} \geq 30^o$), it was obtained that $p_v = 0$. For point B at $v_{ej} \geq 16.4$ km/s and $i_{ej} \geq 45^o$, $p_v$ was greater than for other starting points. For $T = 10$ Myr, the ratio $p_v/p_e$ was mainly less than 1 at $v_{ej} \leq 11.5$ km/s and was greater than 1 at $v_{ej} \geq 14$ km/s. At $v_{ej} = 12$ km/s a result of comparison of $p_v/p_e$ with 1 depended on $i_{ej}$ and the point of ejection.

At $T = 1$ Myr and $i_{ej} = 45^o$, the probabilities $p_v$ equaled to 0.04, 0.02, and 0 for point W at $v_{ej}$ equal to 11.5, 12, and 14–20 km/s, respectively. For other five points of ejection, they were in intervals 0.01–0.05, 0.004–0.08, and 0.0004–0.02, respectively. The ratio of the values of $p_v$ at $T$ to those at $T_{end}$ in Table 6 was in the ranges 0–0.24, 0–0.53, and 0–0.93 at $T$ equal to 0.1, 1, and 10 Myr, respectively. Such ratio can differ much for different ejection velocities and points.

### 3.5. Probabilities of collisions of bodies ejected from the Earth with Mars, Mercury, and Jupiter

Probabilities $p_{me}$ of collisions of bodies ejected from the Earth with Mercury at $T = T_{end}$ were about 0.02–0.08 and 0.03–0.05 at $11.3 \leq v_{ej} \leq 11.5$ and $12 \leq v_{ej} \leq 20$ km/s, respectively (see Fig. 2c and Table 4). At $T = 10$ Myr almost at any $v_{ej}$ and $i_{ej}$, the values of $p_{me}$ were in the intervals 0–0.05 (see Fig. S1c). The only exclusion was for the series *vb* at $i_{ej} \leq 45^o$. In these calculations, $p_{me}$ could exceed 0.12.

The probabilities of collisions of bodies with Mars were smaller than those with Mercury and did not exceed 0.012 and 0.025 at $T = 10$ Myr and $T = T_{end}$, respectively (see Figs. 1d and S1d). They were 0 at $T = 1$ Myr in most calculation variants.

In most calculation variants with 250 bodies, there were no collisions with Jupiter. However, at $T = T_{end}$ in a few variants, mainly with large $v_{ej}$, the probabilities $p_j$ of collisions of bodies with Jupiter exceeded 0.01. They equaled to 0.012, 0.02, and 0.024 in the variants (*vf*, $v_{ej} = 16.4$ km/s, $i_{ej} = 45^o$), (*vw*, $v_{ej} = 16.4$ km/s, $i_{ej} = 45^o$), and (*vw*, $v_{ej} = 14$ km/s, $i_{ej} = 89^o$), respectively. Without any weights to an initial position, an ejection velocity, and an ejection angle, the mean value of $p_j$ over all considered variants equaled to 0.0014. For Mars, similar averaged value of $p_{ma}$ equaled to 0.006, and the fraction of variants with non-zero values of $p_{me}$ was greater than that with non-zero values of $p_{ma}$.

### 3.6. Probabilities of the collisions of bodies ejected from the Earth with the Sun and probabilities of the ejection of bodies into hyperbolic orbits

In the variants *vf, vc, vu,* and *vd* at $T = T_{end}$ and $v_{ej} = 11.3$ km/s, the probability $p_{sun}$ of collisions of bodies with the Sun varied from 0.1 to 0.6 (see Table 4 and Fig. 2e). Often $p_{sun}$ was between 0.2 and 0.5, and the upper limit of $p_{sun}$ was greater at greater $v_{ej}$. For $T = 10$ Myr, it was obtained that $p_{sun} \leq 0.1$ at $11.3 \leq v_{ej} \leq 12$ km/s, and $p_{sun} \leq 0.25$ at $14 \leq v_{ej} \leq 16.4$ km/s. The intervals of $p_{sun}$ for ejection from points B and W differed from those from four other points of ejection at $11.22 \leq v_{ej} \leq 11.25$ km/s and at $v_{ej} \geq 14$ km/s. For $11.22 \leq v_{ej} \leq 11.25$ km/s, $p_{sun}$ could reach 0.08 at the *vb* and *vw* series, and $p_{sun} = 0$ at other series. For $v_{ej} \geq 14$ km/s, depending on $i_{ej}$, the values of $p_{sun}$ for ejection from points B and W could be smaller (be up to 0) or greater (be up to 0.7) than from other points. The ratio of the probability of collisions of bodies with the Sun at $T = T_{end}$ to that at $T = 10$ Myr could reach 5. The ratio of the probability of collisions of bodies with the Earth to the probabilities of collisions of bodies with other planets and the Sun usually decreased with time.

The fraction $p_{ej}$ of bodies ejected from the Solar System was mainly greater for greater ejection velocity. For $T = T_{end}$, the values of $p_{ej}$ did not exceed 0.06 and 0.08 at $v_{ej} \leq 11.4$ km/s and $11.5 \leq v_{ej} \leq 12$ km/s, respectively. For the *vw* variants, $p_{ej}$ equaled to 1 at $v_{ej} = 16.4$ km/s and $i_{ej} \geq 60^o$ and at $v_{ej} = 20$ km/s and $i_{ej} \leq 45^o$. At $i_{ej} = 45^o$ and $v_{ej} = 16.4$ km/s, the fraction of bodies ejected from the Solar System was about 0.25–0.27 if points of ejection were not in the front or back of the Earth's motion. For five points of ejection (other than W) and $v_{ej} \leq 16.4$ km/s, $p_{ej}$ was smaller for greater $i_{ej}$. For many calculation variants at $v_{ej} \leq 12$ km/s, there was no ejection at $T \leq 1$ Myr, and more than a half of ejections were after 10 Myr (see Fig. 3c). In contrast, at $v_{ej} \geq 16.4$ km/s more than a half of ejections could be during the first million years. Usually there were almost no ejections into hyperbolic orbits after 10 Myr at $v_{ej} \geq 16.4$ km/s. Therefore, the plots of $p_{ej}$ via $i_{ej}$ presented in Figs. 2f and S1f are very close at $T = 10$ Myr and $T = T_{end}$ for $v_{ej} \geq 16.4$ km/s. Exclusive for the *vw* series, at $v_{ej} = 20$ km/s the values of $p_{ej}$ were smaller for greater $i_{ej}$. In the variants *vf, vc, vu,* and *vd*, they were about 0.4 at $i_{ej} \leq 30^o$ and $v_{ej} = 20$ km/s. For the W starting point, $p_{ej}$ was about 0.05 at $i_{ej} = 15^o$ and $v_{ej} \geq 16.4$ km/s.

The average values of $p_e$, $p_v$, $p_{me}$, $p_{ma}$, $p_{sun}$, and $p_{ej}$ depend on the distributions of bodies over $v_{ej}$ and $i_{ej}$ and on points of ejection. The obtained values of probabilities of collisions of ejected bodies with planets can be used for more accurate estimates of the probabilities for the model of ejection of bodies which will take into account the distributions of ejected bodies over ejection velocities, ejection angles, and ejection points. Such distributions depend on the parameters of impact collisions which cause such ejections.

### 3.7. Probabilities and velocities of collisions of bodies ejected from the Earth with the Moon

Ipatov (2024a) studied the probabilities of collisions of bodies ejected from the Earth with the Moon in its present orbit and the velocities of collisions of ejected bodies with the Earth and the Moon. In (Ipatov, 2024a), as in the present paper, the Moon was not included in the integration of motion equations of bodies. Based on the arrays of orbital elements of migrated bodies, similar to (Ipatov and Mather, 2003, 2004a, 2004b; Ipatov, 2019), Ipatov (2024a) calculated the probabilities $p_M$* and $p_E$* of collisions of bodies with the Moon and the Earth and the ratio $k_{EM} = p_E$*/$p_M$* of probabilities of collisions of bodies with the Earth to those with the Moon. The probability $p_M$ of a collision of a body with the Moon was calculated as $p_M = p_E/k_{EM}$, where $p_E$ was calculated at integration of motion equations as in the present paper. The ratio $k_{EM}$ was about 35–48, 25–32, and 15–30 at $v_{ej}$ equal to 11.3, 12, and 16.4 km/s, respectively.

The probability of collisions of bodies ejected from the Earth with the Moon was mainly about 0.004–0.008 at ejection velocities $v_{ej} \geq 14$ km/s and about 0.006–0.01 at $v_{ej} = 12$ km/s. It was greater at lower ejection velocities and was in the range of 0.01–0.02 at $v_{ej} = 11.3$ km/s.

In the considered calculations of the motion of bodies ejected from the Earth, the ejected bodies left the Hill sphere of the Earth and moved in heliocentric orbits. With the present orbit of the Moon, the probability of collisions of the ejected bodies with the Moon in its present orbit for the bodies that did not leave the Hill sphere of the Earth was probably even less than that for the bodies that moved in heliocentric orbits. Note that for the model of ejection of a body from the Earth in a random direction, the square of the ratio of the radius of the Moon to the radius $R_{Mo}$ of the Moon's orbit is only $2 \cdot 10^{-5}$. This square of the ratio corresponds to the probability of a collision of a body with the Moon during a single body's cross of the sphere with a radius equal to $R_{Mo}$. A high probability of the collision of a body ejected from the Earth with the Moon could be only if the ejected body moved for a long time inside the Hill sphere of the Earth, and the Moon moved close to the Earth. In my calculations, even for small ejection velocities, the duration of motion of bodies inside the sphere did not exceed 10 years.

As it is noted in many papers (e.g. considering the Giant Impact Model of Moon formation, discussed in Section 1.1), in order to contain the present fraction of iron, the Moon had to accumulate the main fraction of its mass from the mantle of the Earth. The Moon got some material ejected from the Earth during the accumulation of the Earth and during the Late Heavy Bombardment. Based on the obtained probabilities of collisions of bodies ejected from the Earth with the Moon, Ipatov (2024a) noted that the bodies ejected from the Earth and falling onto the Moon embryo would not be enough for the Moon to grow to its present mass from a small embryo moving in the present orbit of the Moon. This result testifies in favour of the formation of a lunar embryo and its further growth to most of the present mass of the Moon near the Earth. Such formation is considered in the Giant Impact and multi-impact models of Moon formation (see references to such papers in Section 1.1). For more efficient growth of the Moon embryo with an initial mass of no more than 0.1 of the present mass of the Moon (e.g., considered in (Ipatov, 2018)), it is desirable (Ipatov, 2024a) that after some collisions of large impactors with the Earth, the ejected bodies should not simply fly out of the crater, but some of the matter should get orbits around the Earth, as in the multi-impact model (e.g., Rufu et al., 2017, 2021). Note that Ipatov (2018) supposed that the embryos of the Earth and the Moon were formed from the common rarefied condensation with a mass of about 0.01–0.1$m_E$. This condensation could get the angular momentum needed for such binary formation at a collision of

| $T$ | $h$, km | $p_{sun}$ | $p_{me}$ | $p_v$ | $p_e$ | $p_{ma}$ | $p_{ej}$ |
|---|---|---|---|---|---|---|---|
| 10, Myr | 0 | 0.068 | 0.02 | 0.184 | 0.148 | | 0.008 |
| 10, Myr | 100 | 0.104 | 0 | 0.136 | 0.132 | 0.004 | 0.012 |
| 10 Myr | 1000 | 0.104 | 0.004 | 0.12 | 0.144 | 0 | 0.016 |
| $T_{end}$ | 0 | 0.372 | 0.08 | 0.32 | 0.196 | 0.008 | 0.02 |
| $T_{end}$ | 100 | 0.372 | 0.048 | 0.3 | 0.236 | 0.008 | 0.036 |
| $T_{end}$ | 1000 | 0.424 | 0.044 | 0.26 | 0.208 | 0.012 | 0.048 |

two rarefied condensations moved in almost circular orbits. Such supposition about formation of the initial Moon embryo helps to understand why other terrestrial planets do not have large satellites. Large bodies collided with all terrestrial planets, and therefore such planets could have large satellites if one considers the multi-impacts model. For the model by Ipatov (2018), the Moon embryo already existed before it began to accumulate material ejected from the Earth. Similar embryos did not form near other terrestrial planets because their parent rarefied condensations had not such angular momentum which was needed for formation of a binary (not a single body) during contraction of a condensation. The angular momentum needed for such contraction was discussed by Nesvorný et al. (2010). The results obtained by Nesvorný et al. (2010) were also used by Ipatov (2017) for his model of formation of trans-Neptunian binaries, which is similar to the model of Moon formation presented in (Ipatov, 2018).

Velocities of collisions of bodies ejected from the Earth with the Moon and the Earth depended on ejection angles $i_{ej}$, ejection velocities $v_{ej}$, and ejection points. The average velocities of collisions of ejected bodies with the Earth are greater at greater ejection velocity. The values of these collision velocities were about 13, 14–15, 14–16, 14–20, and 14–25 km/s at ejection velocities equal to 11.3, 11.5, 12, 14, and 16.4 km/s, respectively. The velocities of collisions of bodies with the Moon were also higher at greater ejection velocities and were mainly in the range of 7–8, 10–12, 10–16, and 11–20 km/s at $v_{ej}$ equal to 11.3, 12, 14, and 16.4 km/s, respectively (Ipatov, 2024a).

### 3.8. Dependence of the orbital evolution of ejected bodies on a height of ejection

Earlier Gladman et al. (2005) and Reyes-Ruiz et al. (2012) made calculations at the height $h$ of initial position of ejected bodies above the Earth's surface equal to 100 km. In order to understand whether such assumption gives results similar to calculations at $h = 0$, and at what $h$ the results are close to those for calculations at $h = 0$, the calculations with different values of $h$ were made. It was concluded that the influence of the height $h \leq 100$ km on the probabilities of collisions of ejected bodies with planets is greater for a smaller ejection velocity $v_{ej}$, and it is considerable only when $v_{ej}$ is close to the escape velocity. For example, for point F, $v_{ej} = 11.22$ km/s, and $i_{ej} = 45^o$, all bodies fell back onto the Earth in 15 days at $h = 0$ or $h = 1$ km, in 25 days at $h = 20$ km, and in 21 years at $h = 50$ km. In this case at $h = 100$ km, 235 bodies of 250 initial bodies left the Hill sphere of the Earth and in total 130 bodies (52 %) fell back onto the Earth.

For ejection from point F, $v_{ej} = 12$ km/s, and $i_{ej} = 45^o$ at $T = 10$ Myr or $T = T_{end}$ and at three values (0, 100, and 1000 km) of a height $h$ of ejection, the fractions $p_{sun}$, $p_{me}$, $p_v$, $p_e$, and $p_{ma}$ of bodies collided with the Sun, Mercury, Venus, Earth, and Mars and the fraction $p_{ej}$ of ejected bodies are presented in Table 7. The results obtained at $h = 20$ km were close to those at $h = 100$ km.

The values of the probabilities presented in Table 7 at $h = 0$ differed a little from the values at greater values of $h \leq 1000$ km. It is possible to see in Table 7 slightly larger values of $p_{ej}$ and smaller values of $p_{me}$ at $h \geq 100$ km than at $h = 0$.

The differences in the probabilities of collisions of ejected bodies with planets at different $h$ do not differ much from the differences between similar probabilities at the same $h$, but at slightly different initial velocities, also for other calculation variants. For example, for point F, $v_{ej} = 16.4$ km/s, $i_{ej} = 45°$, and $T = T_{end}$ at four values of $h$ equal to 0, 1, 10, and 100 km, the probabilities $p_{me}$ equaled to 0.1, 0.05, 0.08, and 0.06, the values of $p_v$ equaled to 0.19, 0.24, 0.25, and 0.24, and $p_e$ equaled to 0.10, 0.15, 0.11, and 0.12, respectively.

For bodies started from $h$ equal to 3, 5, or 7 radii of the Earth (such distances correspond to the Moon embryo moving close to the Earth) at the *vf* series (at $v_{ej} = 12$ km/s and $i_{ej} = 45°$), the values of $p_{ej}$, $p_v$, and $p_e$ at $T = T_{end}$ were about 0.16–0.2, 0.23–0.26, and 0.13–0.14, respectively, i. e. for such values of $h$, $p_e$ was smaller and $p_{ej}$ was greater than at $h = 0$.

## 4. The amount of material ejected from the Earth that could be delivered to planets

### 4.1. Ejection of material from the Earth due to its collisions with bodies from its feeding zone

In order to understand the amount of ejected material delivered to planets at different time, we need to estimate the amount of material delivered to the Earth and then ejected from the Earth outside its Hill sphere. For giant impactors, the ejection of material depended much on parameters of impacts. Galimov (2011) believed that the material ejected during the giant impact should consist of 80–90 % vapor. Though at giant impacts a lot of material was ejected in the form of gas and dust and a disk was formed around the Earth, some fraction of ejected material could leave the Hill sphere of the Earth. Analyzing the data from Table 1 in (Cuk and Stewart, 2012), we can conclude that for the collision of an impactor with the almost formed Earth at an impact velocity $v_{imp} = 20$ km/s for a mass $m_{imp}$ of an impactor equal to $0.026m_E$ and $0.1m_E$, the ratio $(\Delta m_E + m_{imp}{-}m_{disk})/m_{imp}$ equaled 1.4 and 1.3, respectively, where $\Delta m_E$ is the variation in the mass of the Earth's embryo after the impact, $m_{disk}$ is the mass of a disk of ejected material (from which the Moon embryo formed). Therefore, at such giant impacts the total mass $m_H$ of material that left the Hill sphere of the Earth could exceed the mass of an impactor, i.e. could be up to the mass of Mars. For example, at the total mass of bodies that left the Hill sphere of the Earth equal to $0.02m_E$ and at the fraction of bodies collided with Venus equal to $p_v = 0.2$, the total mass of material ejected from the Earth and collided with Venus could be equal to $0.004m_E$. Such exchange of material between the Earth and Venus could be at a giant impact. The probabilities of collisions of material ejected after giant impacts with planets were discussed by Emsenhuber et al. (2021) and are shortly presented in Section 1.3.

Based on the results of calculations presented in Section 3, below in Section 4 I discuss the model of ejection of bodies at the impacts for which all ejected material was in the form of bodies. Such model probably is for impactors with diameters about 1–100 km. For small impactors, ejected material did not get velocities greater than the escape velocity. So the below estimates for such model show the upper limit of material ejected from the Earth outside its Hill sphere.

The amount of material ejected from the Earth that was delivered to a planet is the production $p_{pl}·m_{ej}$ of the amount $m_{ej}$ of material that was ejected from the Earth and of the probability $p_{pl}$ of collisions of bodies with the planet. Such probabilities were discussed above in Section 3. At the late stages of planet formation, masses and orbits of planets could differ from the present values. However, the main conclusions based of the calculated probabilities of collisions of bodies ejected from the Earth with planets can be similar to those for a more accurate model, as I made only approximate estimates of the delivery of ejected bodies to planets. The values of $m_{ej}$ are proportional to the mass $m_{imp}$ of bodies delivered to the Earth from considered distances from the Sun. As it is discussed in Section 1.2, the ratio $k_{ej} = m_{ej}/m_{imp}$ typically does not exceed 0.2 and is different for different collisions. The amounts $m_{imp}$ of material delivered to the Earth from several distances from the Sun at different times are discussed below.

Simulations by Woo et al. (2022) yield ~$0.25m_E$ of leftover planetesimals that can contribute to late accretion. Earth analogues typically finish accreting over 90 % of their mass before 80 Myr of the simulation. During the final growth of the mass of the Earth when it collided with bodies with a total mass $m_{imp}$ equal to $0.1m_E$, the Earth collided mainly with planetesimals from the whole feeding zone of the terrestrial planets. The value of ejected mass $m_{ej}$ could be about $0.01m_E$ at $m_{imp} = 0.1m_E$ and $k_{ej} = m_{ej}/m_{imp} = 0.1$. Characteristic velocities $v_{imp}$ of collisions of planetesimals with the Earth were about 13–19 km/s (Marov and Ipatov, 2021). Armstrong et al. (2002) noted that ejection velocities are generally less than $0.85v_{imp}$. For $13 < v_{imp} < 19$ km/s, the values of $0.85v_{imp}$ are about 11–16 km/s. For such ejection velocities, we can estimate $p_e$ to be about 0.2–0.3. At $p_e = 0.3$ and $m_{ej} = 0.01m_E$, the total mass $m_{ej}p_e$ of such ejected bodies that could fall back onto the Earth could be about $0.003m_E$. At $m_{ej} = 0.01m_E$ for $p_{Moon} = 0.006$, $p_v = 0.3$, $p_{me} = 0.06$, $p_{ma} = 0.008$, and $m_{ej} = 0.01m_E$, estimates of the total mass of bodies ejected from the Earth and collided with the Moon, Venus, Mercury, and Mars could be of the order of $0.00006m_E$, $0.003m_E$, $0.0006m_E$, and $0.00008m_E$, respectively. Similar estimates of such masses can be made for other values of $m_{ej}$ presented below, taking into account that the above probabilities and the ratio $m_{ej}/m_{imp}$ can be different for different characteristic impact and ejection velocities.

### 4.2. Ejection of material from the Earth due to its collisions with the bodies that came from beyond Mars's orbit

A lot of material was delivered to the Earth from beyond Mars's orbit. Joiret et al. (2023) calculated formation of the terrestrial planets for the Nice model, including the motion of planetesimals from the ring between 21 and 30 AU with a total mass equal to $25m_E$. They concluded that the total mass of cometary bodies from such disk delivered to the Earth could be between $7 \times 10^{-7}m_E$ and $8 \times 10^{-4}m_E$.

Ipatov (2020, 2021, 2024b), Ipatov and Mather (2003, 2004a, 2004b), Marov and Ipatov (2018), Marov and Ipatov (2023) studied the motion of bodies under the gravitational influence of planets moving in their present orbits. Ipatov and Mather (2003, 2004a, 2004b) showed that the probability of a collision with the Earth for a Jupiter-crossing comet was about $4 \times 10^{-6}$ - $4 \times 10^{-5}$. The range was for different groups of such comets. It was noted that though a typical lifetime of such Jupiter-crossing objects (JCOs) is about 0.1 Myr, a few JCOs among considered 30,000 JCOs after some time got orbits with aphelia inside Jupiter's orbit and could move in such orbits for tens of million years. The probability of a collision with the Earth for the former JCO which got an orbit with such aphelion distance could be greater than the total probability of thousands other JCOs with similar initial orbits. For initial orbits of JCOs with semi-major axes between 4.5 and 12 AU and initial eccentricity $e_o = 0.3$, the probability of their collisions with the Earth was about $2 \times 10^{-6}$ (Marov and Ipatov, 2018). Our above estimates of the fractions $p_E$ of bodies collided with the Earth are not differed much from estimates by Morbidelli et al. (2000), who obtained the probability of a collision of a particle with the Earth to be of the order of $10^{-6}$-$3·10^{-6}$ per particles in the 5–8 AU region, decreasing to $5·10^{-7}$ for larger semi-major axes (8–9 AU). Their probabilities are a little smaller than in our calculations because Morbidelli et al. (2000) considered circular initial orbits, and in our calculations initial orbits were eccentric. In my opinion, due to gravitational influence of planets and planetesimals, eccentric initial orbits could better match real starting orbits of considered bodies.

The orbital evolution of bodies from disks with a width equal to 2.5 AU located at a distance from 5 to 40 AU from the Sun (with $e_o = 0.05$ or $e_o = 0.3$) was studied in (Ipatov, 2020), and from disks with a width equal to 0.1 AU located at a distance from the Sun from 3 to 5 AU (with $e_o = 0.02$ or $e_o = 0.15$) it was studied in (Ipatov, 2021). The values of the probability $p_E$ of a collision of one of bodies with initial semi-major axes between 5 and 40 AU from the Sun with the Earth are mainly between $10^{-6}$ and $10^{-5}$ (Ipatov, 2020, 2024b; Marov and Ipatov, 2023).

For some initial semi-major axes of orbits of bodies between 3.2 and 3.6 AU, the probability $p_E$ of a collision of a body with the Earth could exceed 0.001 and even 0.01 (Ipatov, 2021; Marov and Ipatov, 2023). Some bodies that migrated from such distances could collide with the Earth after a few hundred million years. During variation of the semi-major axis of the orbit of Jupiter, the region of initial semi-major axes of bodies with such values of $p_E$ could be wider. Ejection of planetesimals into hyperbolic orbits caused the decrease in the semi-major axis of the orbit of Jupiter. Results of computer simulations of the evolution of disks of planetesimals with the giant planets showed (Ipatov, 1987) that such decrease was about $0.003M_g/m_E$ AU, where $M_g$ is the total mass of planetesimals in the zone of the giant planets. In (Ipatov, 1993) such mass was considered to be 135-150$m_E$ and variations in the semi-major axis of Jupiter's orbit were 0.005$M_g/m_E$ AU. At $M_g = 100m_E$ the semi-major axis of Jupiter's orbit could decrease by 0.5 AU. Therefore, for such evolution of Jupiter's orbit, for the zone of initial semi-major axes of orbits of bodies for which the probability of a collision with the Earth could be up to 0.001–0.01, instead of the mentioned above zone located at 3.2–3.6 AU from the Sun, we can consider the wider zone 3.2–4.1 AU. All resonances in the asteroid belt were also shifted during the decrease of the semi-major axis of Jupiter's orbit.

Other models of variation of the semi-major axis of Jupiter's orbit were considered in the Nice models (e.g., Clement et al. 2018, 2019; Gomes et al., 2005; Morbidelli et al., 2005, 2010; Tsiganis et al., 2005) and in the Grand Tack model (Jacobson and Morbidelli, 2014; O'Brien et al., 2014; Rubie et al., 2015; Walsh et al., 2011). In (Gomes et al., 2005) the semi-major axes of embryos of Uranus and Neptune increased slowly from less than 14 AU from the Sun to less than 18 AU during 880 Myr. Then Jupiter and Saturn crossed the 1:2 mean motion resonance, and the semi-major axes of Uranus and Neptune jumped to their present values. Note that in calculations by Ipatov (1991, 1993) the embryos of Uranus and Neptune moved from initial distances from the Sun less than 10 AU to their present orbits under the gravitational influence of planetesimals with a total mass of about 100$m_E$, and the main growth of the semi-major axes was during 10 Myr. So, Uranus and Neptune could get their present orbits long before Jupiter and Saturn crossed the above resonance. Morbidelli et al. (2007) considered that jumps in semi-major axes of orbits of Uranus and Neptune took place at $T = 2$ Myr and were caused by the crossing of the 3:5 mean motion resonance between Jupiter and Saturn. In the Grand Tack model, Jupiter was first moving, under the presence of gas, toward the Sun, to 1.5 AU. Such motion was before the final gas-free stages of accumulation of the terrestrial planets. Morbidelli et al. (2010) considered the evolution of orbits of asteroids during a sharp change in Jupiter's orbit that led to a sharp change in the positions of resonances. They found that the probability of collisions of asteroids with the Earth was about $8 \times 10^{-4}$. Morbidelli (2013) supposed that about 10–20 % of the mass of the terrestrial planets was accreted from the outer part of the asteroid belt.

Based on the estimates of the total mass $m_\sum$ of bodies that were at some distances from the Sun and on the probabilities $p_E$ of their collisions with the Earth, we can estimate the corresponding total mass of planetesimals delivered to the Earth as $m_{imp} = p_E \cdot m_\sum$. If we consider the probability of collisions of planetesimals for initial distances from the Sun of about 5–10 AU with the Earth to be about $4 \times 10^{-6}$ (Ipatov, 2020; Marov and Ipatov, 2023) and the total mass $m_\sum$ of such planetesimals to be about 100$m_E$, then $m_{imp} = 4 \times 10^{-4}m_E$. For the zone of 10–40 AU at $p_E = 1.5 \times 10^{-6}$ и $m_\sum = 100m_E$, we have $m_{imp} = 1.5 \times 10^{-4}m_E$. However, above estimates are probably majorant estimates of $m_{imp}$, as only a part of initial number of planetesimals was left in the zone of the giant planets up to the late stages of formation of the terrestrial planets. Similar to the above estimates of $m_\sum$, Safronov (1972) considered the total mass of bodies initially located in the zone of the giant planets to be of the order of 100$m_E$. Fernandez and Ip (1984) made calculations for formed Jupiter and Saturn, embryos of Uranus and Saturn, with initial masses equal to 0.2 of their present masses, and a disk of planetesimals with initial distances from the Sun between 12 and 40 AU and with the total mass of the disk up to 20 present masses of Uranus and Neptune (i. e., up to 600$m_E$). Calculations by Ip (1989) were made for already formed Jupiter and Saturn and a disk of planetesimals with initial distances from the Sun between 10 and 20 AU and with a total mass equaled to 100$m_E$. For the Nice model, Gomes et al. (2005) considered the mass of the disk of planetesimals between 15.5 and 34 AU from the Sun to be equal to 35$m_E$, and Morbidelli et al. (2007) made calculations for similar disks with masses equal to 50$m_E$ and 65$m_E$.

For the zone of about 3–4 AU from the Sun at $p_e = 10^{-3}$ (Ipatov, 2021; Marov and Ipatov, 2023) and $m_\sum = 10m_E$, we have $m_{imp} = 0.01m_E$. Note that the outer asteroid belt and the feeding zone of the giant planets each could deliver to the Earth the amount of water which is in the Earth's oceans. Though the zone of the outer asteroid belt contained less material with less fraction of water than the zone of the giant planets, it could deliver more water to the Earth. As it is discussed above, the zone of the outer asteroid belt with $p_E$ about $10^{-3}$ earlier was greater than now due to the variation of the semi-major axis of Jupiter during formation of the Solar System. A few other sources of the Late Heavy Bombardment are discussed in Introduction.

The fraction $k_{ej} = m_{ej}/m_{imp}$ of bodies ejected outside the Hill sphere of the Earth typically did not exceed 0.2, and it was greater for greater velocities of collision of impactors with the Earth. Typical velocities of ejection from the Earth are greater for greater collision velocities. As it is seen from Fig. 2, for greater ejection velocity the probabilities of collisions of ejected bodies are typically smaller for collisions with the Earth, and are greater for collisions with Mercury. Characteristic velocities of bodes migrated from the zone of the outer asteroid belt and from the feeding zone of the giant planets and then collided with the Earth were about 23–26 km/s (Marov and Ipatov, 2021). At such collision velocities, ejection velocities of some bodies ejected from the Earth could exceed 20 km/s (see estimates at the end of Section 3.7). As it is noted above in the paper, the probability $p_e$ of a collision of an ejected body with the Earth is smaller at greater collision velocities. Because of different collision and ejection velocities for different sources of incoming impactors, the value of $p_e$ could be less for the bodies migrated from outside Mars's orbit than for the bodies migrated from the feeding zone of the terrestrial planets. Based on the data presented in Tables 3 and 4 and in Fig. 2, we can conclude that such difference in probabilities for different sources probably is smaller for probabilities of collisions of ejected bodies with other terrestrial planets than with the Earth. For bodies that came from outside the orbit of Mars, characteristic velocities of collisions with the Earth were greater, and so the values of $p_v$ could be a little smaller, and the values of $p_{me}$ and $p_j$ could be greater than for bodies from the terrestrial feeding zone. The values of $p_{ma}$ are mainly less than 0.01 in both cases. As for the above text, $p_v$, $p_{me}$, $p_{ma}$, and $p_j$ are probabilities of collisions of bodies ejected from the Earth with Venus, Mercury, Mars and Jupiter, respectively. Probably, in the case of collisions of bodies migrated from beyond Mars's orbit, we can estimate $p_e$, $p_v$, $p_{me}$, and $p_{ma}$ to be about 0.2, 0.2, 0.08, and 0.008, respectively. The total mass of bodies ejected from the Earth and then collided with a planet can be estimated as $m_{imp} \cdot k_{ej} \cdot p_{pl}$, using the above values of $m_{imp}$ and $p_{pl}$, where $p_{pl}$ is a probability for the considered planet, $k_{ej}$ is not more than 0.2, $m_{imp}$ is the total mass of impactors collided the Earth that came from one of the regions considered above.

Some material could be ejected from the Earth in the form of dust, and dust could be produced at destruction of ejected bodies. Note that the probabilities of collisions of dust particles with planets are typically greater than for bodies (Ipatov, 2010). Dust could be more effective than bodies in the delivery of organic matter to planets because dust particles would not have experienced as significant heating as larger bodies during the descent in the atmosphere. In our studies (Ipatov and Mather, 2006; Ipatov et al., 2004; Ipatov, 2010) of the motion of dust particles originated in different parts of the Solar System, we took into account the gravitational influence of all planets, the Poynting-Robertson drag, radiation pressure, and solar wind drag. Sizes of particles varied from 1 μm to a few millimeters, and the largest probabilities of collisions of particles with the terrestrial planets were for particles with diameters of about 100 μm. Particles of such sizes could be more effective than other dust particles in delivery material to the terrestrial planets.

### 4.3. Recent ejection of bodies from the Earth

According to Nesvorny et al. (2023), about 20 bodies with a diameter $d > 10$ km were expected to hit the Earth between 2.5 and 3.5 Gyr ago. Nesvorny et al. (2021a) stated that there were about 16–32 impacts of near-Earth asteroids with a diameter $d > 5$ km and 2–4 impacts with $d > 10$ km on the Earth in the last 1 Gyr. Craters were formed as the result of collisions of migrated bodies with the terrestrial planets and the Moon. Diameters of the largest lunar craters (Apollo, Hertzsprung) exceed 500 km. More old craters were formed on the far side of the Moon (Head et al., 2010; Neumann et al., 2015). For vertical collisions, Ipatov et al. (2020) considered the dependence of crater diameter $D_v$ on a diameter $D_p$ of an impactor (in km) and a velocity $U$ of a collision (in km/s) as $D_v = 3.94 \, D_p^{0.92} \cdot U^{0.52}$. They obtained that 15 km lunar craters were made by 1 km impactors. The number of 15 km lunar craters formed during the last billion years is about 50 (Mazrouei et al., 2019; Ipatov et al., 2020). The ratio of probabilities of collisions of near-Earth objects with the Earth and the Moon was estimated in (Ipatov et al., 2020) to be equal to 22. Therefore, the number of 1 km bodies that hit the Earth during the last billion years can be estimated to be about 1000, i.e. about 1 hit during 1 Myr. Based on analysis of lunar craters, Ipatov et al. (2020) estimated the mean number of near-Earth objects during the last 1 Gyr. They concluded that the obtained results do not contradict the growth in the number of near-Earth objects and the frequency of their collisions with the Moon by 2.6 times after the probable catastrophic fragmentation of large main-belt asteroids, which according to (Mazrouei et al., 2019) may have occurred 290 Myr ago. So, during last 290 Myr the number of such impacts with the Earth could be even more than on average during last 1 Gyr. The mass of a 1 km impactor can be about $10^9$ tons. Therefore, up to the order of $10^8$ tons of material could be ejected from the Earth at each collision with a 1 km impactor. To estimate the total mass of bodies delivered from the Earth to planets we can multiply this ejected mass by the probabilities of collisions of bodies with planets discussed in Section 4.1.

The ejection of material from the Earth due to an impact event depends on several factors, including the size and velocity of the impactor and the angle between the velocity and the surface, as well as on the geological characteristics of the impact site. Typically, impactors with a diameter greater than 1 km are required for significant ejecta to reach the escape velocity (11.2 km/s). It is unlikely that ejected material reaches the escape velocity for impactors with diameters not more than several tens of meters. As it is discussed in Introduction, most of collisions of bodies with planets were during the first billion years since formation of the Solar System. However, existence of craters shows that bodies with a diameter greater than 1 km collided with the Earth after the LHB. It is considered (e.g., Davis, 2008; Schulte et al., 2010; Allen et al., 2022) that the impactors with a diameter $d_i = 10$–15 km formed the Chicxulub crater (with a diameter $D_{cr} = 150$–200 km and an age $T_{cr} = 65.5$ Myr), Vredefort crater (with $D_{cr} = 170$–300 km and $T_{cr} = 2.02$ Gyr), and Sudbury crater (with $D_{cr} = 130$ km and $T_{cr} = 1.85$ Gyr). Allen et al. (2022) supposed that $d_i$ could be 20–25 km for the Vredefort crater. The Popigai crater (Paquay et al., 2014; Schmitz et al., 2015) is fourth largest verified impact crater on the Earth (with $D_{cr} = 90$–100 km and $T_{cr} = 35.7$ Myr). It could be produced by an impactor with $d_i = 8$ km (Paquay et al., 2014). It is supposed that impactors that produced the above craters came from the asteroid belt. The mass of a 10 km impactor is about the total mass of a thousand 1 km impactors, whose role in ejection of material from the Earth is discussed in the previous paragraph. Bodies ejected from the craters after traveling for a long time in space can be found on the Earth, other terrestrial planets, and the Moon. As dynamical lifetimes of bodies ejected from the Earth can reach hundreds of million years, a small fraction of bodies ejected at the Chicxulub and Popigai events can still move in the zone of the terrestrial planets and have small chances to collide with planets, including the Earth.

Bodies ejected from the Earth could deliver organic material to other celestial objects, e.g. to Mars. Marov (2023) concluded that Mars had previously experienced favorable climatic conditions with an abundance of water on its surface, as indicated by residual geological landforms. He wrote that these conditions dramatically changed around 3.8–3.6 Gyr ago, leaving it with a waterless, barren surface and a sparse atmosphere. Nevertheless, evidence for the existence of a favorable paleoclimate on Mars supports the concept of a habitable zone in the Solar System extending up to the orbit of Mars. Bradak (2023) discussed the possible microorganism containing material transport between the early Earth and Mars and supposed that early Mars, similar to the Earth, might allow biological evolution and might be able to harbor life. So, bodies ejected from the Earth could play some role in the existence of some forms of life on Mars. Venus is too hot for life similar to that on the Earth.

Gladman et al. (2005) and Reyes-Ruiz et al. (2012) considered motion of bodies ejected from the Earth during 30 Kyr because they supposed that during this time ejecta bodies could make the biological seeding of Mars and Venus. The amounts of material that could be ejected at different times from the Earth outside its Hill sphere are discussed in Sections 4.1–4.2. The amounts of this material delivered to Mars and Venus during hundreds of million years could be about 0.01 and 0.2 of the ejected material. In (Reyes-Ruiz et al., 2012) fractions of bodies collided during 30 kyr with Venus and Mars were of the order $10^{-3}$ and $10^{-5}$, respectively. In our Fig. 3a the fraction of bodies collided during about 30 kyr with Venus was about 0.005. For ejection point F, this fraction varied from 0.02 at $v_{ej} = 11.3$ km/s to 0 at $v_{ej} = 16.4$ km/s. The fraction $p_{me}$ of bodies collided with Mercury during 30 kyr depended on a point of ejection. For example, in Fig. 3b at $T = 30$ kyr the value of $p_{me}$ was 0 for point F and can be up to 0.02 for points D and U. For variants presented in this figure, there were no collisions of bodies with Mars during 30 kyr. The above estimates show that in contrast to other terrestrial planets, bodies ejected from the Earth had much less chances to reach Mars in 30 kyr.

Microbes can survive in space for a longer time than 30 kyr. For example, Mileikowsky et al. (2000) wrote that in studying natural transfer of microbes between Mars and Earth, the purpose was to arrive at the order of magnitude of viable arrivals after flight times shorter than 1 million years. When presenting the results in their tables VIIc–XI (the tables considered the number bacteria traveled between the Earth and Mars), flight times of up to 100 kyr and 330 kyr were chosen. Mileikowsky et al. (2000) concluded that the size of the ejected material needs to be bigger than 0.2 m in diameter to protect the microbes from heat during the impact and the escape phase. In my calculations presented in Fig. 3b, the results for 6000 ejected bodies are considered. Only one such body collided with Mars in 1 Myr (at 0.46 Myr after ejection from point W). For these 6000 bodies, the fraction of their collisions with Mars was 0.00017 during 1 Myr. This value is close to the results by Mileikowsky et al. (2000) who concluded that of ejecta emitted from Earth, roughly 0.016 % land on Mars, within about 1 Myr (see their Fig. 2 and the formula above this figure).

## 5. Conclusions

During formation of the Earth and at the stage of the Late Heavy Bombardment, some massive bodies collided with the Earth. Such collisions caused ejection of material from the Earth. The motion of bodies ejected from the Earth was studied, and the probabilities of collisions of the bodies with the terrestrial planets were calculated. The dependences of these probabilities on velocities, angles, and points of ejection of bodies were studied.

Some bodies ejected from the Earth could move in the zone of the terrestrial planets for up to a few hundred million years. On average, about a half and less than 10 % of initial ejected bodies still moved in elliptical orbits after 10 and 100 Myr, respectively. A few bodies collided with planets after 250 Myr.

The fraction of bodies ejected from the Earth that after some time collided again with the Earth is greater for a smaller ejection velocity $v_{ej}$. At $v_{ej} \leq 11.25$ km/s, most of the ejected bodies fell back onto the Earth. Over the considered time interval, at an ejection velocity $v_{ej}$ equal to 11.5, 12, and 14 km/s, the values of the fraction $p_e$ of bodies collided with the Earth were approximately 0.3, 0.2, and 0.15–0.2, respectively. At $v_{ej} = 20$ km/s, the values of $p_e$ equaled to 0 for ejection from the forward point of the Earth's motion (from apex) at an ejection angle $i_{ej}$ not less than 30°, exceeded 0.2 for ejection from the back point (from antapex), and were about 0.1 for ejection from other four considered points of ejection.

The ratio of the values of $p_e$ at the end of evolution to those at time $T = 10$ Myr was usually not more than 2. The fraction $p_e$ of bodies collided with the Earth during the first million years at an ejection angle $i_{ej} = 45°$ for all six considered points of ejection was about 0.4–0.6, 0.07–0.1, 0.008–0.04, and 0.004–0.02 at $v_{ej}$ equal to 11.3, 12, 14, and 16.4 km/s, respectively.

In considered calculations, the fraction $p_v$ of bodies collided with Venus over the entire considered time interval was often about 0.2–0.3. The values of $p_v$ were about $p_e$ at $v_{ej} = 11.5$ km/s, were often greater than $p_e$ at $v_{ej} \geq 12$ km/s, and were less than $p_e$ at $v_{ej} \leq 11.4$ km/s. The total number of ejected bodies delivered to the Earth and Venus probably did not differ much.

The fractions of bodies collided with Mercury over the entire considered time interval (at $T = T_{end}$) were about 0.02–0.08 and 0.03–0.05 at $11.3 \leq v_{ej} \leq 11.5$ and $12 \leq v_{ej} \leq 20$ km/s, respectively. The fractions of bodies collided with Mars were smaller than those with Mercury and did not exceed 0.012 and 0.025 at $T = 10$ Myr and $T = T_{end}$, respectively. More material ejected from the Earth was delivered to Mercury than to Mars. In most calculation variants, there were no collisions with Jupiter. However, at $T = T_{end}$ in a few variants, mainly with large $v_{ej}$, the fraction $p_j$ of bodies collided with Jupiter exceeded 0.01. On average the fraction of bodies collided with Jupiter can be estimated to be of the order of 0.001. The fraction of bodies collided with the Sun often was between 0.2 and 0.5 at different ejection velocities $v_{ej} \geq 11.3$ km/s.

The fraction of bodies ejected into hyperbolic orbits was mainly greater for a greater ejection velocity. Over a whole considered time interval, it did not exceed 0.1 at an ejection velocity not more than 12 km/s. At an ejection velocity greater than 16 km/s and an ejection angle not less than 60 degrees, all bodies were ejected from the Solar System if they started from the front (in the direction of the motion) point of the Earth. At an ejection angle equal to 45° and an ejection velocity equal to 16.4 km/s, the fraction of bodies ejected from the Solar System was about 0.25–0.27 if points of ejection were not in the front or back of the Earth's motion.

The obtained probabilities of collisions of bodies ejected from the Earth with planets can differ for different points, velocities and angles of ejection. More accurate estimates of the probabilities can be made with the use of the models that consider distribution of ejected material over ejection angles, velocities, and points.

The probabilities of collisions of bodies ejected from the Earth with planets could be similar to the probabilities of collisions of planetesimals with planets for planetesimals that were left in the feeding zone of the Earth at the late stages of its formation. The obtained results testify in favour of the hypothesis that the upper layers of the Earth and Venus can contain similar material. Smaller fractions of material ejected from the Earth can be found on other terrestrial planets and Jupiter.

Bodies ejected from the Earth could deliver organic material to other celestial objects, e.g. to Mars.

As dynamical lifetimes of bodies ejected from the Earth can reach hundreds of million years, a small fraction of bodies ejected at the Chicxulub and Popigai events can still move in the zone of the terrestrial planets and have small chances to collide with planets, including the Earth.

## CRediT authorship contribution statement

**S.I. Ipatov:** Writing – review & editing, Formal analysis, Conceptualization.

## Declaration of competing interest

There is no conflict of interests involved in the present study.

## Data availability

Data will be made available on request.

## Acknowledgments

This work was supported by the Ministry of Science and Higher Education of the Russian Federation, within the budget theme of the Vernadsky Institute of Geochemistry and Analytical Chemistry of Russian Academy of Sciences. I thank the two reviewers for their helpful comments and suggestions.

## Appendix A. Supplementary data

Supplementary data to this article can be found online at https://doi.org/10.1016/j.icarus.2024.116341.

**Probabilities of collisions of bodies ejected from forming Earth with the terrestrial planets**
S.I. Ipatov

The fractions of bodies that collided with the terrestrial planets or the Sun or ejected into hyperbolic orbits during the first 10 Myr.

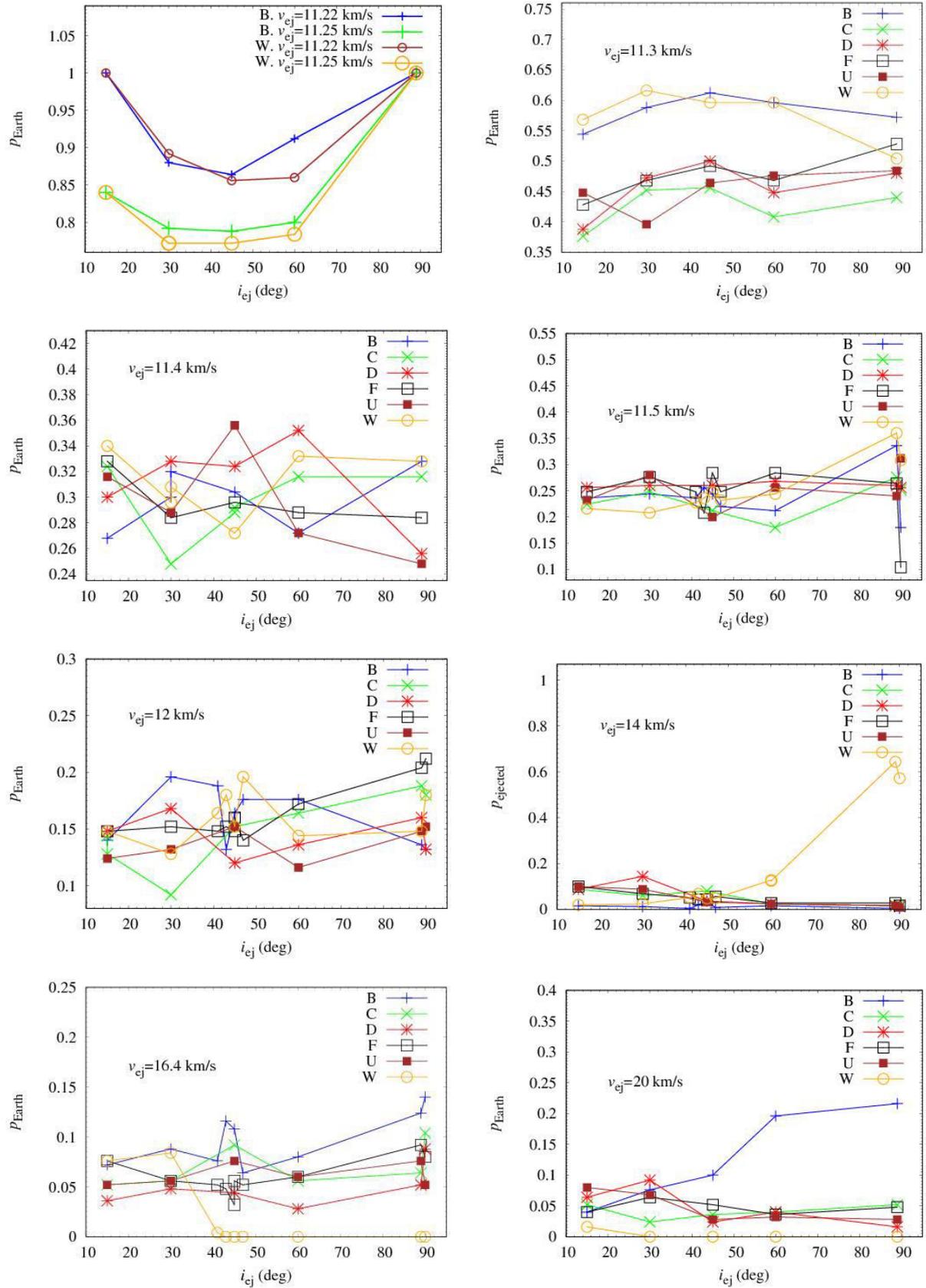

Fig. S1a

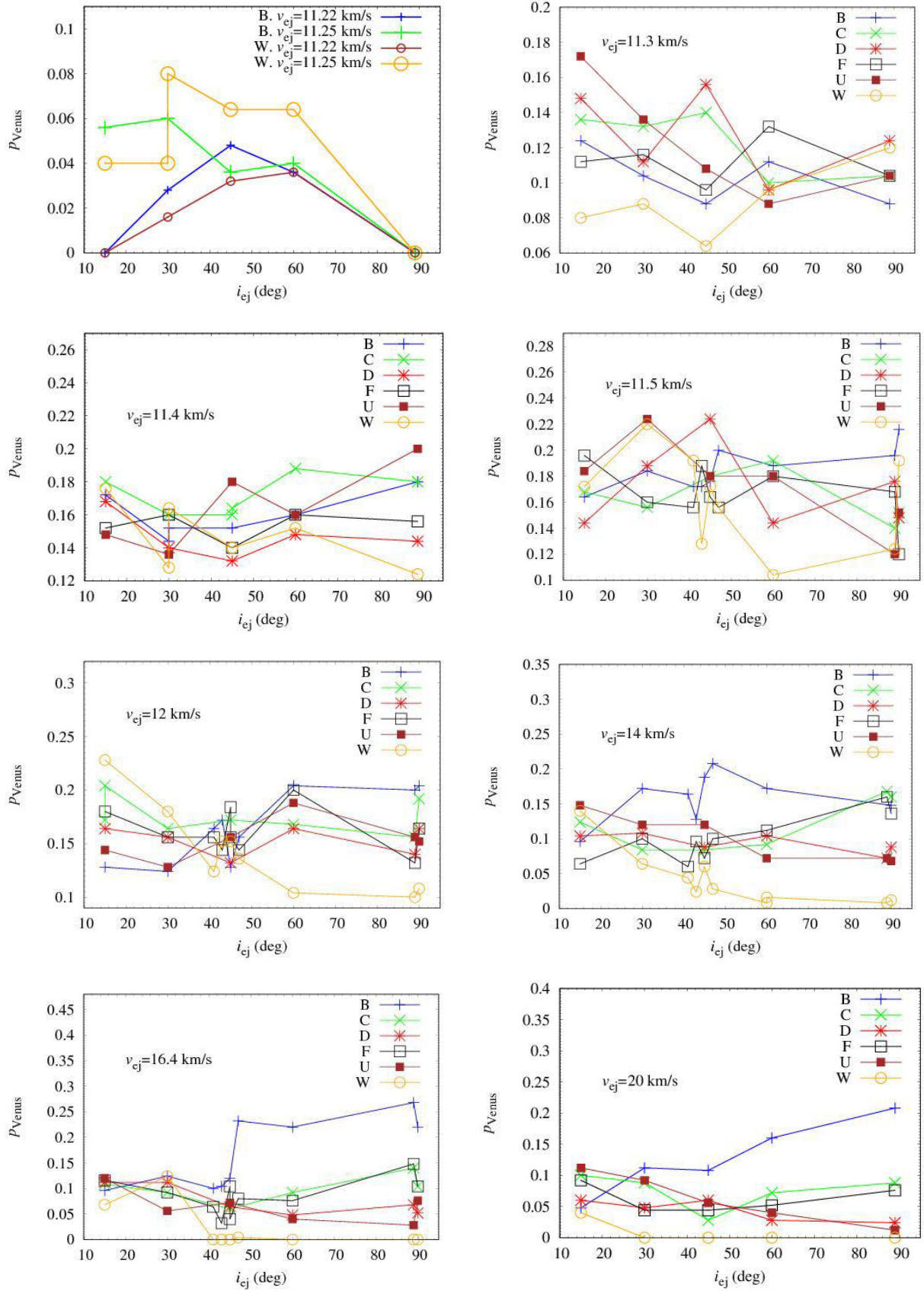

Fig. S1b

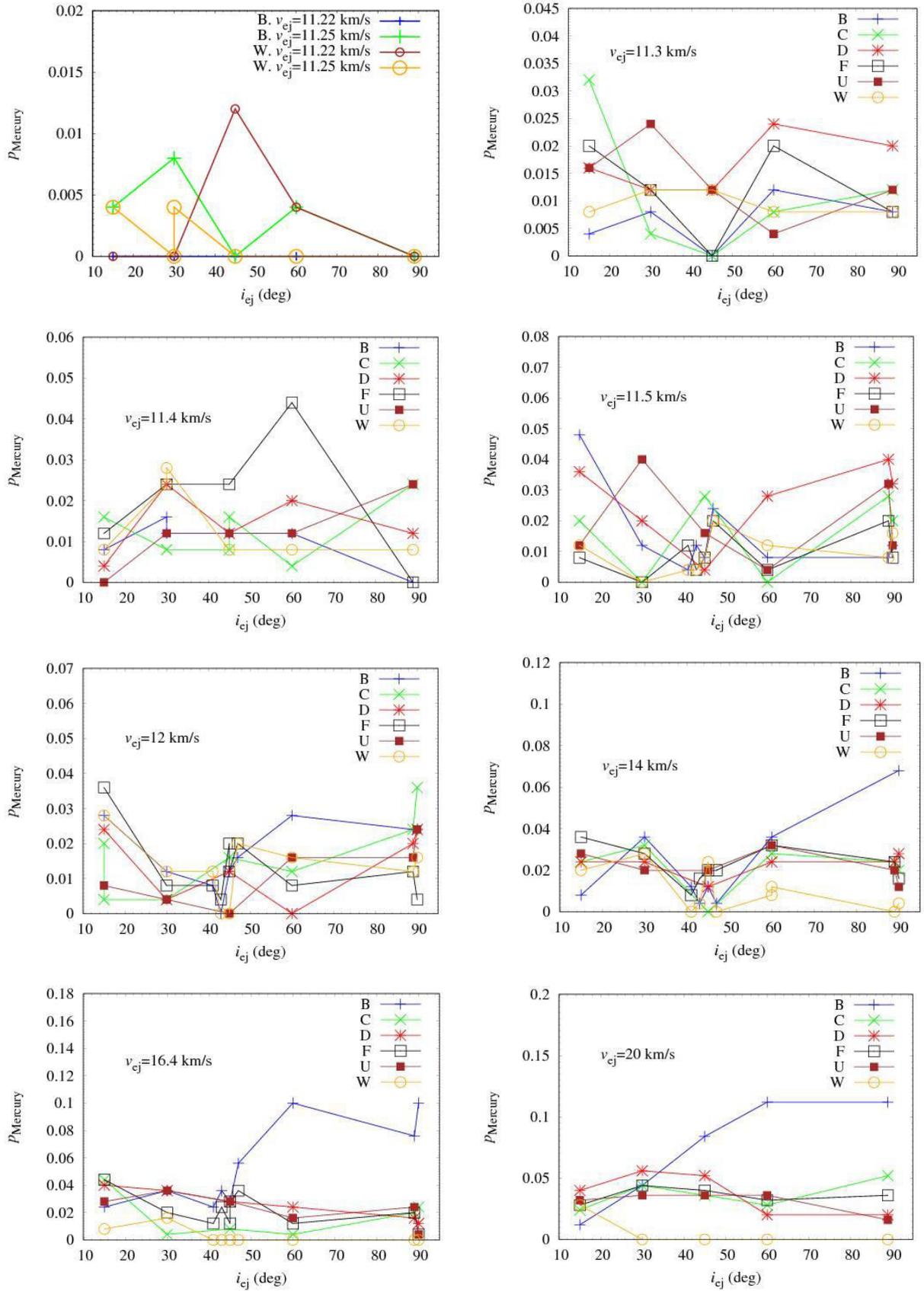

Fig. S1c

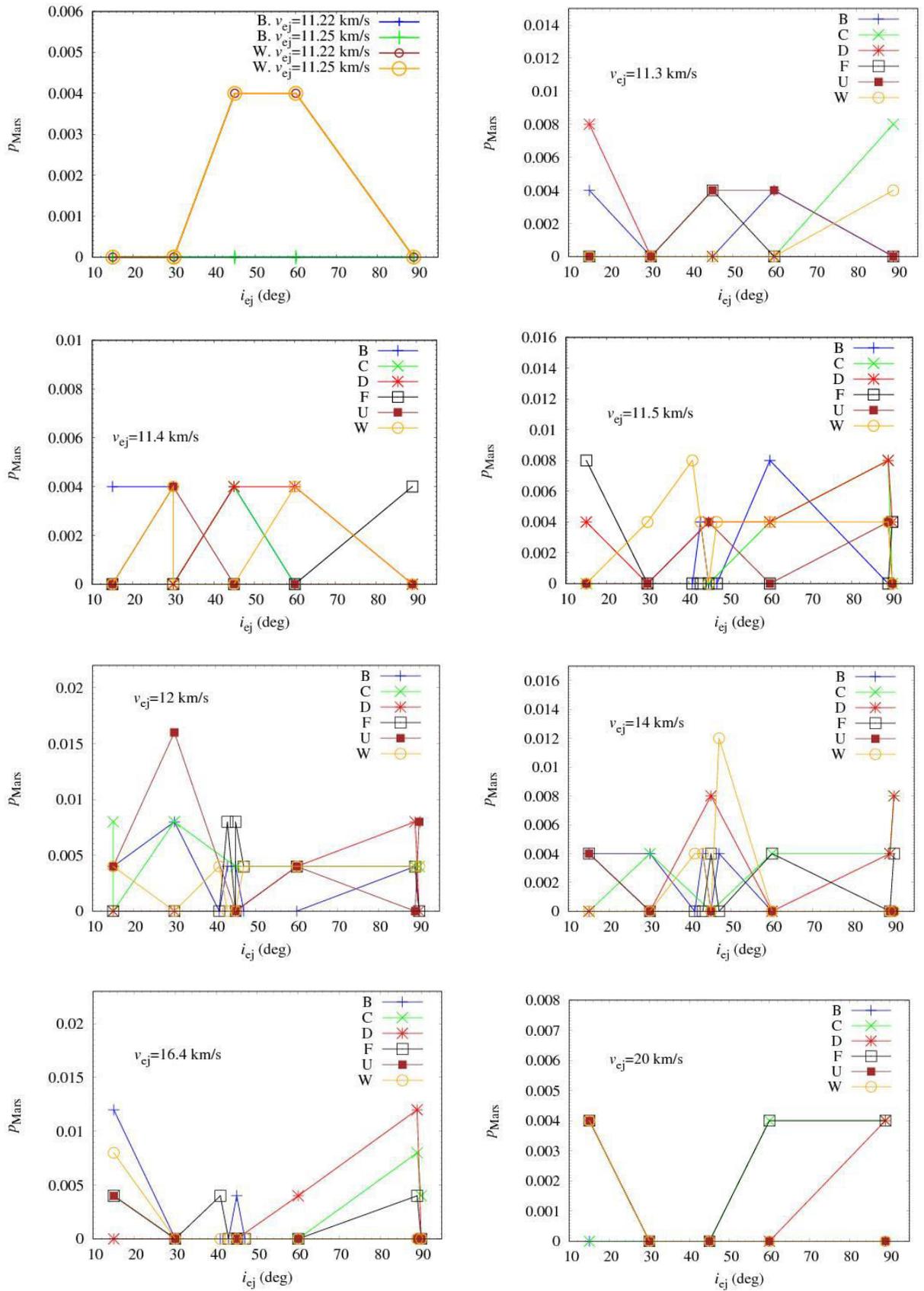

Fig. S1d

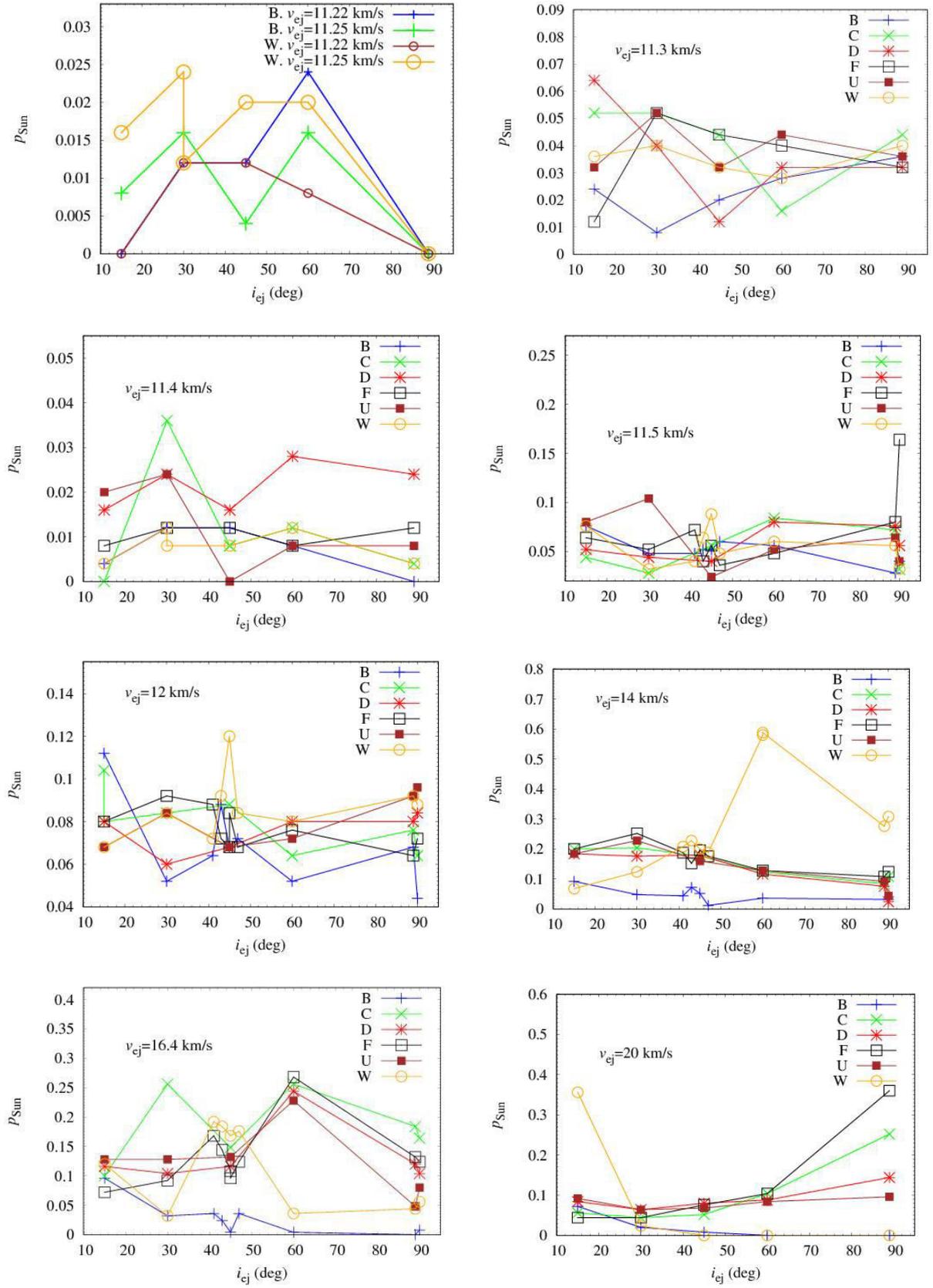

Fig. S1e

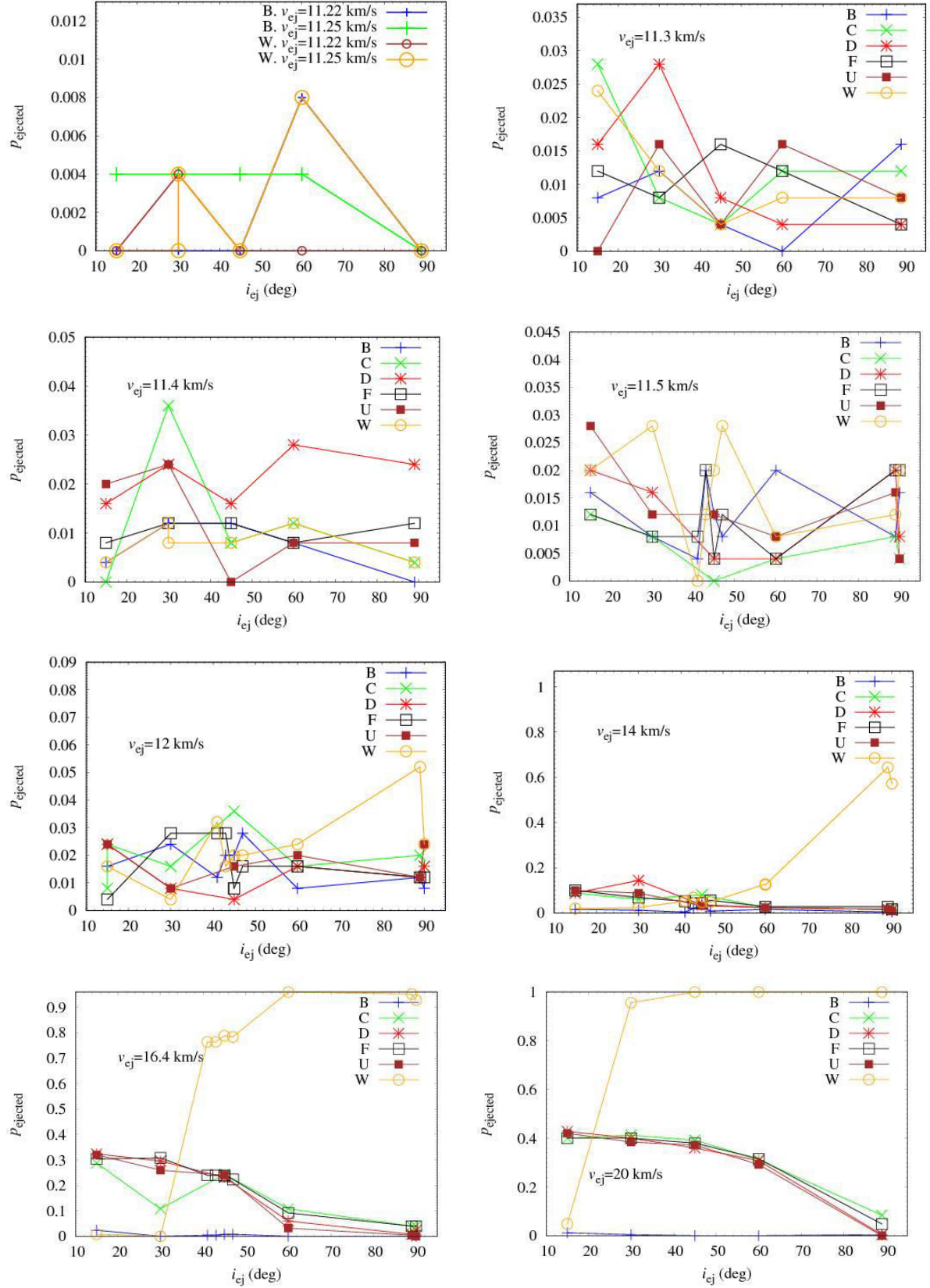

Fig. S1f

**Fig. S1.** The fractions of bodies that collided with the Earth ($p_e$, Fig. S1a), Venus ($p_v$, Fig. S1b), Mercury ($p_{me}$, Fig. S1c), Mars ($p_{ma}$, Fig. S1d), the Sun ($p_{sun}$, Fig. S1e) or ejected into hyperbolic orbits ($p_{ej}$, Fig. S1f) during the first 10 Myr vs. the ejection angle $i_{ej}$ at several values of an ejection velocity $v_{ej}$ and six points of ejection (B, C, D, F, U, and W). The values of the fractions are presented for $i_{ej}+i_{ts}$, where $i_{ts}$ is equal to -4°, -2°, 0, or 2° for $t_s$ equal to $1^d$, $2^d$, $5^d$, or $10^d$, respectively. Each sign on the figure corresponds to the mean value for 250 bodies.